\documentclass[a4paper,fleqn,usenatbib]{mnras}

\usepackage{newtxtext,newtxmath}

\usepackage[T1]{fontenc}
\usepackage{ae,aecompl}

\usepackage{graphicx}	
\usepackage{amsmath}	
\usepackage{amssymb}	

\usepackage{enumitem}
\setlist[itemize]{leftmargin=*}

\usepackage{epsfig}
\usepackage{multirow}
\usepackage{float}
\usepackage{footmisc}
\def\mnras{MNRAS}

\def\apjl{ApJL}
\def\apjs{ApJS}
\def\aap{A\&A}

\def\nat{Nature}

\def\xmm{{\it XMM-Newton}}

\def\rxj04{RX J0439.6-5311}
\def\rej1034{RE J1034+396}
\def\1h07{1H 0707-495}
\def\pg12{PG 1244+026}
\def\grs1915{GRS 1915+105}

\title[Re-observing \rej1034. I. the Recurrent X-ray QPO]{Re-observing the NLS1 Galaxy \rej1034. I. the Long-term, Recurrent X-ray QPO with a High Significance}

\author[C. Jin, et al.]{
Chichuan Jin$^{1,2}$\thanks{E-mail: ccjin@nao.cas.cn},
Chris Done$^{3}$,
Martin Ward$^{3}$
\\
$^{1}$National Astronomical Observatories, Chinese Academy of Sciences, 20A Datun Road, Beijing 100101, China\\
$^{2}$School of Astronomy and Space Sciences, University of Chinese Academy of Sciences, 19A Yuquan Road, Beijing 100049, China\\
$^{3}$Centre for Extragalactic Astronomy, Department of Physics, University of Durham, South Road, Durham DH1 3LE, UK\\
}

\date{prepared for MNRAS}

\pubyear{2019}

\begin{document}
\label{firstpage}
\pagerange{\pageref{firstpage}--\pageref{lastpage}}
\maketitle

\begin{abstract}
\rej1034\ is a narrow-line Seyfert 1 galaxy (NLS1) in which the first significant X-ray quasi-periodic oscillation (QPO) in an active galactic nuclei (AGN) was observed in 2007. We report the detection of this QPO in a recent \xmm\ observation in 2018 with an even higher significance. The quality factor of this QPO is 20, and its period is $3550\pm80$ s, which is $250\pm100$ s shorter than in 2007. While the QPO's period has no significant energy dependence, its fractional 
root-mean-square (rms) variability increases from 4\% in 0.3-1 keV to 12\% in 1-4 keV bands. An interesting phenomenon is that the  QPO in 0.3-1 keV {\it leads} that in the 1-4 keV bands by $430\pm50$~s with a high coherence, opposite to the soft X-ray {\it lag} reported for the observation in 2007. We speculate that the QPO has an intrinsic hard lag, while the previous reported soft lag is caused by the interference of stochastic variability. This soft X-ray lead in the new data supports the idea that the QPO of \rej1034\ is a possible AGN counterpart of the 67~Hz high-frequency QPO seen in the black hole binary (BHB) GRS 1915+105. We also search for QPO harmonics, but do not find any significant signals. Our new data reinforce previous results that the QPO is seen in a specific spectral state, as the only 2 observations showing no significant QPO signal exhibit an even stronger soft X-ray excess than the other 6 observations which display the QPO. Therefore, our results imply that the QPO in \rej1034\ is physically linked to a soft X-ray component.
\end{abstract}




\section{Introduction}
\subsection{X-ray QPO in AGN}
Both stellar mass black hole binaries (BHB) and active
galactic nuclei (AGN) are powered by gas accreting onto a central black hole, and 
their observational properties are determined primarily by the black hole mass, mass
accretion rate and spin. This relative simplicity should allow us to scale their observed X-ray properties such as variability and spectra between these two very different black hole mass systems. However, while the broadband power spectral densities (PSD) do show some similarities
(M$^{c}$Hardy et al. 2006, 2007), the BHB show 
strong quasi-periodic oscillations (QPOs) at both low frequencies
(0.1-10~Hz: potentially from Lense-Thirring precession: Stella \& Vietri 1998; Ingram, Done \& Fragile 2009; Veledina, Poutanen \& Ingram 2013) and high frequencies (100s of Hz: potentially related to the Keplerian period of the innermost disc: Remillard \& McClintock 2006), which are generally absent in AGN.

The lack of QPO detections in AGN is probably mainly due to the much longer timescale of AGN QPO predicted by scaling their from BHB. Even the lowest-mass AGN of $\sim 10^6 M_\odot$ would have predicted mass-scaled low-frequency QPOs at 0.1-10 day timescales, which makes them difficult to study with the restricted duration of continuous X-ray exposures (Vaughan \& Uttley 2005, 2006). More typical local AGN with masses of 
$\sim 10^{7-8} M_\odot$ would imply much worse data windowing problems. Instead, high-frequency QPOs provide a better potential match to observational constraints for the lowest-mass AGN. These are observed locally 
as Narrow-line Seyfert 1 galaxies (NLS1s), which are accreting at high Eddington ratios (e.g. Done \& Jin 2016; Jin, Done \& Ward 2016, 2017a,b). Indeed, the first AGN X-ray QPO was discovered in the NLS1 \rej1034\ with a period of $3730\pm60$ s (Geirli\'{n}ski et al. 2008). Since then a few X-ray QPOs with lower significances have been reported in NLS1s, such as 1H 0707-495 (Pan et al. 2016; Zhang et al. 2018), MS 2254.9-3712 (Alston et al. 2015), Mrk 766 (Zhang et al. 2017), MCG-06-30-15 (Gupta et al. 2018). A couple of Seyfert 2s were also reported to exhibit a QPO, including 2XMM J123103.2+110648 (Lin et al. 2013) and XMMU J134736+173403 (Carpano \& Jin, 2018). X-ray QPOs were reported in tidal disruption events around super-massive black holes, such as Swift J164449.3+573451 (Reis et al. 2012) and ASASSN-14li (Pasham et al. 2019). Recently, a new type of X-ray periodic signal given the term quasi-periodic eruption (QPE) has been reported in the Seyfert 2 galaxy GSN 069, whose black hole mass is estimated to be $\sim 4\times10^{5}M_{\odot}$ (Shu et al. 2018; Miniutti et al. 2019), although the properties of X-ray QPE are very different from QPO.

\begin{table*}
 \centering
   \caption{List of \xmm\ Observations on \rej1034. GTI is the integrated good exposure time in EPIC-pn after removing intervals containing background flares. Obs-1 and Obs-2 are in the full-frame mode, while the other observations are all in the small-window mode. $N_{\rm H,host}$ is the best-fit host galaxy absorption (see Section~\ref{sec-spec-sx}), and the Galactic absorption is fixed at $1.36\times10^{20}$cm$^{-2}$. $F_{\rm 0.3-2 keV}$ is the absorbed 0.3-2 keV flux. Errors indicate the 90\% confidence range. $f_{\rm QPO}$ and $Q_{\rm QPO}$ are the QPO frequency and quality factor in the 1-4 keV band as reported by Alston et al. (2014) for the first 8 observations. The data of Obs-1 is not good enough for the QPO analysis due to severe background contamination. Obs-3 and Obs-6 are the two observations when the QPO is not detected. Obs-9 is our new observation.}
\begin{tabular}{@{}lccccccccccc@{}}
\hline
Obs No. & ObsID & Obs Date & On-Time & GTI & $N_{\rm H,host}$ & $F_{\rm 0.3-2 keV}$ & $F_{\rm 2-10 keV}$ & $\Gamma_{\rm 0.3-2 keV}$  & $\Gamma_{\rm 2-10 keV}$ & $f_{\rm QPO}$ & $Q_{\rm QPO}$ \\
& & & (ksec) & (ksec) & ($10^{20}$cm$^{-2}$) &\multicolumn{2}{c}{(10$^{-12}$ erg cm$^{-2}$ s$^{-1}$)} & & & ($10^{-4}$Hz) & \\
\hline
Obs-1 & 0109070101 & 2002-05-01 & 12.8 & 1.8 & 0.00$^{+0.58}_{l}$ & 8.72$^{+0.29}_{-0.29}$ & 0.95$^{+0.39}_{-0.39}$ & 3.86$^{+0.05}_{-0.05}$ & 1.91$^{+0.37}_{-0.27}$ & -- & -- \\
Obs-2 & 0506440101 & 2007-05-31 & 91.1 & 79.5 & 1.00$^{+0.16}_{-0.16}$ & 8.52$^{+0.04}_{-0.04}$ & 1.16$^{+0.04}_{-0.04}$ & 3.81$^{+0.01}_{-0.01}$ & 2.06$^{+0.06}_{-0.06}$ & 2.7 & 24 \\
Obs-3 & 0561580201 & 2009-05-31 & 60.8 & 43.4 & 1.08$^{+0.11}_{-0.11}$ & 11.27$^{+0.04}_{-0.04}$ & 0.82$^{+0.03}_{-0.03}$ & 4.20$^{+0.01}_{-0.01}$ & 2.09$^{+0.05}_{-0.05}$ & $\times$ & $\times$ \\
Obs-4 & 0655310101 & 2010-05-09 & 44.3 & 19.3 & 0.00$^{+0.17}_{l}$ & 7.97$^{+0.05}_{-0.05}$ & 1.07$^{+0.05}_{-0.05}$ & 3.73$^{+0.01}_{-0.01}$ & 2.03$^{+0.06}_{-0.06}$ & 2.7 & 11 \\
Obs-5 & 0655310201 & 2010-05-11 & 53.0 & 31.2 & 0.00$^{+0.18}_{l}$ & 7.92$^{+0.04}_{-0.04}$ & 1.13$^{+0.04}_{-0.04}$ & 3.71$^{+0.01}_{-0.01}$ & 1.97$^{+0.05}_{-0.05}$ & 2.5  & 13 \\
Obs-6 & 0675440301 & 2011-05-07 & 32.2 & 18.2 & 2.29$^{+0.16}_{-0.16}$ & 13.56$^{+0.07}_{-0.07}$ & 1.02$^{+0.05}_{-0.05}$ & 4.40$^{+0.01}_{-0.01}$ & 1.96$^{+0.06}_{-0.06}$ & $\times$ & $\times$ \\
Obs-7 & 0675440101 & 2011-05-27 & 36.0 & 14.7 & 0.01$^{+0.26}_{-0.01}$ & 8.91$^{+0.07}_{-0.07}$ & 1.20$^{+0.07}_{-0.07}$ & 3.86$^{+0.01}_{-0.01}$ & 1.97$^{+0.07}_{-0.07}$ & 2.6 & 9 \\
Obs-8 & 0675440201 & 2011-05-31 & 29.4 & 12.6 & 0.04$^{+0.27}_{-0.04}$ & 8.12$^{+0.07}_{-0.07}$ & 1.24$^{+0.07}_{-0.07}$ & 3.73$^{+0.01}_{-0.01}$ & 1.87$^{+0.07}_{-0.07}$ & 2.6 & 7 \\
Obs-9 & 0824030101 & 2018-10-30 & 71.6 & 64.7 & 0.00$^{+0.02}_{l}$ & 7.99$^{+0.03}_{-0.03}$ & 1.09$^{+0.03}_{-0.03}$ & 3.73$^{+0.01}_{-0.01}$ & 2.01$^{+0.05}_{-0.05}$ & 2.8 & 20 \\
\hline
\end{tabular}
\label{tab-obs}
\end{table*}

\subsection{\rej1034}
\rej1034\ is a well studied AGN located at $z$ = 0.042. It has an extraordinary steep soft X-ray spectrum compared to other AGN (Puchnarewicz et al. 1995; Wang \& Netzer 2003; Casebeer et al. 2006; Crummy et al. 2006) though much of this is probably due to the disc itself (Done et al. 2012; Jin et al. 2012a,b,c). The hydrogen Balmer emission lines of \rej1034\ have a full width half maximum (FWHM) of $\lesssim 1500$ km s$^{-1}$, defining the source as a NLS1 galaxy (Puchnarewicz et al. 1995; Mason, Puchnarewicz \& Jones 1996; Gon\c{c}alves, V\'{e}ron \& V\'{e}ron-Cetty 1999; Bian \& Huang 2010). Its black hole mass is estimated to be $10^6-10^7~M_{\odot}$ (see Czerny et al. 2016 for a summary of several  different mass estimates), with the most probable mass range being $(1-4)\times10^{6}~M_{\odot}$ (Gerli\'{n}ski et al. 2008; Middleton et al. 2009; Bian \& Huang 2010; Jin et al. 2012a; Chaudhury et al. 2018). The mass accretion rate of \rej1034\ is close to or slightly above the Eddington limit (Jin et al. 2012a; Czerny et al. 2016).

The most notable phenomenon of \rej1034\ is the QPO signal detected in its X-ray emission, which is the first significant detection of an X-ray QPO in AGN (Gierli\'{n}ski et al. 2008). Since then many studies have been conducted in order to understand the physical origin of this QPO, as well as its potential trigger. The QPO varies significantly in its root-mean-square (RMS) amplitude between different observations, but not in its frequency. The QPO signal was most significant in the first detection during the \xmm\ observation in 2007 (Gierli\'{n}ski et al. 2008; Middleton et al. 2009). Then it was detected in only 4 of the 6 subsequent \xmm\ observations made before 2011 (Alston et al. 2014). The high coherence of this QPO signal ($Q\gtrsim10$) is comparable to the high-frequency QPO at 67 Hz seen in the high mass accretion rate state of the BHB GRS 1915+105 (M$~=12.4^{+2.0}_{-1.8}~M_{\odot}$, Reid et al. 2014). This is also consistent with the mass scaling if the mass of \rej1034\ is $(1-4)\times10^{6}~M_{\odot}$ (Middleton, Uttley \& Done 2011; Czerny et al. 2016; Chaudhury et al. 2018).

The RMS of the QPO is energy dependent, showing that the QPO spectrum is subtly different to the time averaged spectrum, and the hard X-ray QPO 
leads the soft X-ray by 300-400 s (Gierli\'{n}ski et al. 2008; Middleton, Done \& Uttley 2011). This corresponds to a light travel distance of $\sim$30 $R_{\rm g}$ in the disc reprocessing scenario, which however places no constraints on the black hole spin. This soft X-ray lag was also reported by Zoghbi \& Fabian (2011) who performed spectral-timing analysis in the frequency domain using the same dataset.

\subsection{This Work}
Despite all previous studies, the long-term behaviour (over a timescale of 10 years) of the QPO in \rej1034\ remains unknown. This is 
because of the visibility issue of this source with \xmm\ since 2011. In this paper, we present results from our new \xmm\ observation of \rej1034\ obtained in 2018. These new data allow us to explore the latest properties of this QPO signal, and help us to understand the mechanism of AGN QPO in general.

This paper is organized as follows. In Section 2 we list all the \xmm\ observations of \rej1034\ and describe the data reduction procedures. In Section 3 we present the light curve and QPO signal in the new data, which is followed by a detailed analysis and modelling of the PSD and QPO in Section 4. The study of the QPO's long-term variation is presented in Section 5. Detailed discussions of the QPO mechanism is presented in Section 6, and the final section summarizes our main results and conclusions. Unless otherwise specified, all the error bars presented in this work refer to the 1$\sigma$ uncertainty.

\section{Observations and Data Reduction}
\label{sec-obs}
\rej1034\ was previously observed by \xmm\ (Jansen et al. 2001) for 8 times between 2002 and 2011, after which it was no longer observed by \xmm\ due to restricted visibility, and so the QPO signal could not be monitored. Since 2018 the visibility has improved to $\gtrsim$ 70 ks per \xmm\ orbit, and so we observed it again with \xmm\ in 2018 for 72 ks in order to reexamine its X-ray QPO. This new observation is already 7 years from the previous observation in 2011, and 11 years from the initial discovery of QPO in 2007. All of the observations are listed in Table~\ref{tab-obs}.

We downloaded all the data from \xmm\ Science Archive (XSA). In this study, we mainly focused on the X-ray variability and QPO, so only the data from the European Photon Imaging Cameras (EPIC) (Str\"{u}der et al. 2001) were used. The \xmm\ Science Analysis System (SAS v18.0.0) was used to reduce the data. Firstly, the {\sc epproc} and {\sc emproc} tasks were used to reprocess the data with the latest calibration files. Then we defined a circular region with a radius of 80 arcsec centered on the position of \rej1034\ as the source extraction region. In the first two observations the EPIC cameras were in the full-window mode, so the background extraction region was chosen to be the same size in a nearby region without any sources. Later observations were all taken in the small-window mode, so for the two Metal Oxide Semi-conductor (MOS) cameras we extracted the background from a nearby Charge-Coupled Device (CCD) chip, while for the pn camera the background was extracted close to the edge of the small window to minimize contamination of the primary source. We adopted good events (FLAG=0) with PATTERN $\leq$ 4 for pn and PATTERN $\leq$ 12 for MOS1 and MOS2.

The {\sc evselect} task was used to extract the source and background light curves, where the background flares were identified. By running the {\sc epatplot} task, we found that the first two observations in the full-window model suffered from significant photon pile-up in the central $\sim$10 arcsec region of the point spread function (PSF), while the following observations were not affected by this effect thanks to the small-window mode used. The {\sc epiclccorr} task was used to perform the background subtraction, and apply various corrections to produce the intrinsic source light curve. The source and background spectra were also extracted using the {\sc evselect} task. Then the {\sc arfgen}, {\sc rmfgen} and {\sc grppha} tasks were run to produce the auxiliary and response files and rebin the spectra. The {\sc Xspec} software (v12.10.1, Arnaud 1996) was used to perform all the spectral analysis. All the timing results presented in this paper were based on the EPIC-pn data with the highest signal-to-noise (S/N) among the three EPIC cameras. The MOS data were reduced in a similar way and used for the consistency check.

\begin{figure}
\centering
\includegraphics[trim=0.15in 0.3in 0.0in 0.0in, clip=1, scale=0.49]{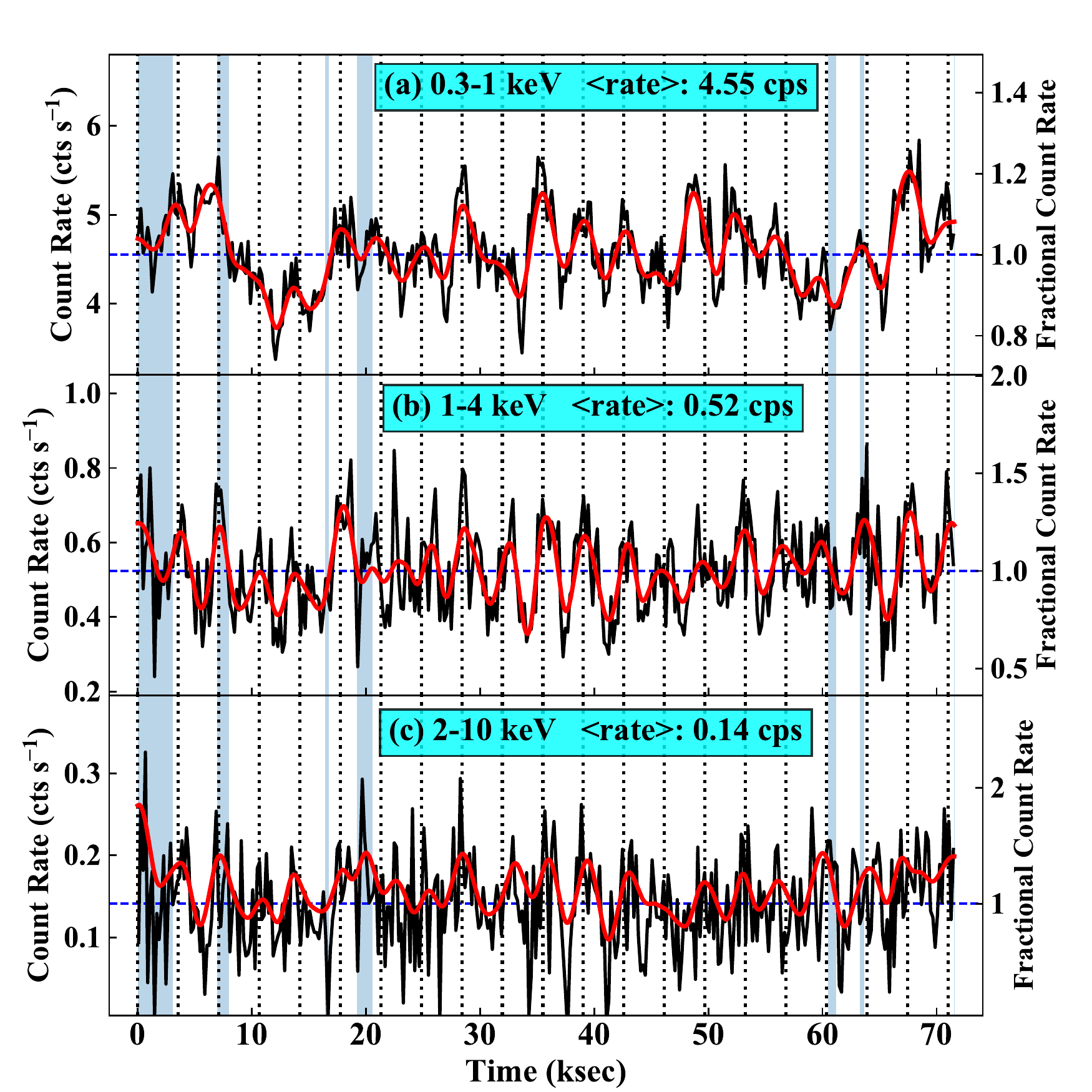} 
\caption{Light curves of \rej1034\ in Obs-9 as observed by \xmm\ EPIC-pn, binned with 200 s. In each panel, the shadowed regions indicate masked time intervals due to background flares. The red solid line is the summation of IMFs whose timescales are equal or longer than the QPO period (see Section~\ref{sec-lc}). The vertical dotted lines indicate every 3550 s time interval, which is the latest period of the 0.3-10 keV light curve.}
\label{fig-lc}
\end{figure}

\section{The New \xmm\ Observation in 2018}
\label{sec-obs9}
We first explore the X-ray variability of \rej1034\ during the latest \xmm\ observation, and search for a QPO signal in the light curve.
\subsection{X-ray Light-curves}
\label{sec-lc}
\rej1034\ exhibits significant X-ray variability in the latest \xmm\ observation in 2018 (hereafter: Obs-9), as shown by the EPIC-pn light curves in Figure~\ref{fig-lc}. The shadowed regions in the figure indicate time intervals affected by background flares. For $\sim$90 per cent of the observing time the background was very low and stable, only the first $\sim$ 4 ks and a few short periods are affected by flares, so the overall data quality is excellent. After masking out all of the background flares, the mean source count rates are found to be 4.55, 0.52 and 0.14 counts per second (cps) in the three typical energy bands of 0.3-1, 1-4 and 2-10 keV, respectively. This immediately suggests that the X-ray spectrum of \rej1034\ remained soft during the new observation. These energy bands are representative because the 0.3-1 keV band is dominated by the soft excess, the 2-10 keV band is dominated by the hard X-ray corona emission (e.g. Middleton et al. 2009). The 1-4 keV band is chosen to facilitate comparison with previous studies, because the QPO signal was significantly detected in this band in 5 out of all 8 \xmm\ observations before 2011 (Alston et al. 2014). In Figure~\ref{fig-lc}, from the y-axis of fractional count rate relative to the mean value, it is also clear that the amplitude of the hard X-ray variability is much larger than that in the soft X-ray band, but the soft X-rays seem to have stronger variability over long timescales ($>$1ks) than short timescales ($<$1ks).

We apply the Ensemble Empirical Mode Decomposition (EEMD) method (Huang et al. 1998; Wu \& Huang 2009; Hu etal. 2011) to these light curves in order to examine the variability in different timescales. This method works in the time domain to resolve a noisy light curve into a complete set of independent components, namely the Intrinsic Mode Functions (IMFs), which possess different variability patterns and are locally orthogonal to each other. This method has been previously applied to the light curve of \rej1034\ for Obs-2, and the QPO variability is found to concentrate in one of the IMFs (Hu et al. 2014).

The Python {\sc PyEMD} package was used to perform the EEMD analysis. We find that each light curve (50 s binned) can be decomposed into 9 IMFs, with the timescale increasing from the first component (C1) to the last (C9). We also find that the QPO signal is contained in C5, while C6 to C9 can be combined to show the variability over longer timescales. The summation of IMFs whose timescales are equal or longer than the QPO period is shown in Figure~\ref{fig-lc} as the solid red line. The periodic positions separated by 3550 s (see Section~\ref{sec-qpovar-freq}) are marked by the vertical dotted lines. It is clearly seen that the instantaneous period of the QPO is varying within the observing time, confirming that it is indeed a quasi-periodic signal.

\begin{figure}
\centering
\includegraphics[trim=0.0in 0.3in 0.0in -0.2in, clip=1, scale=0.59]{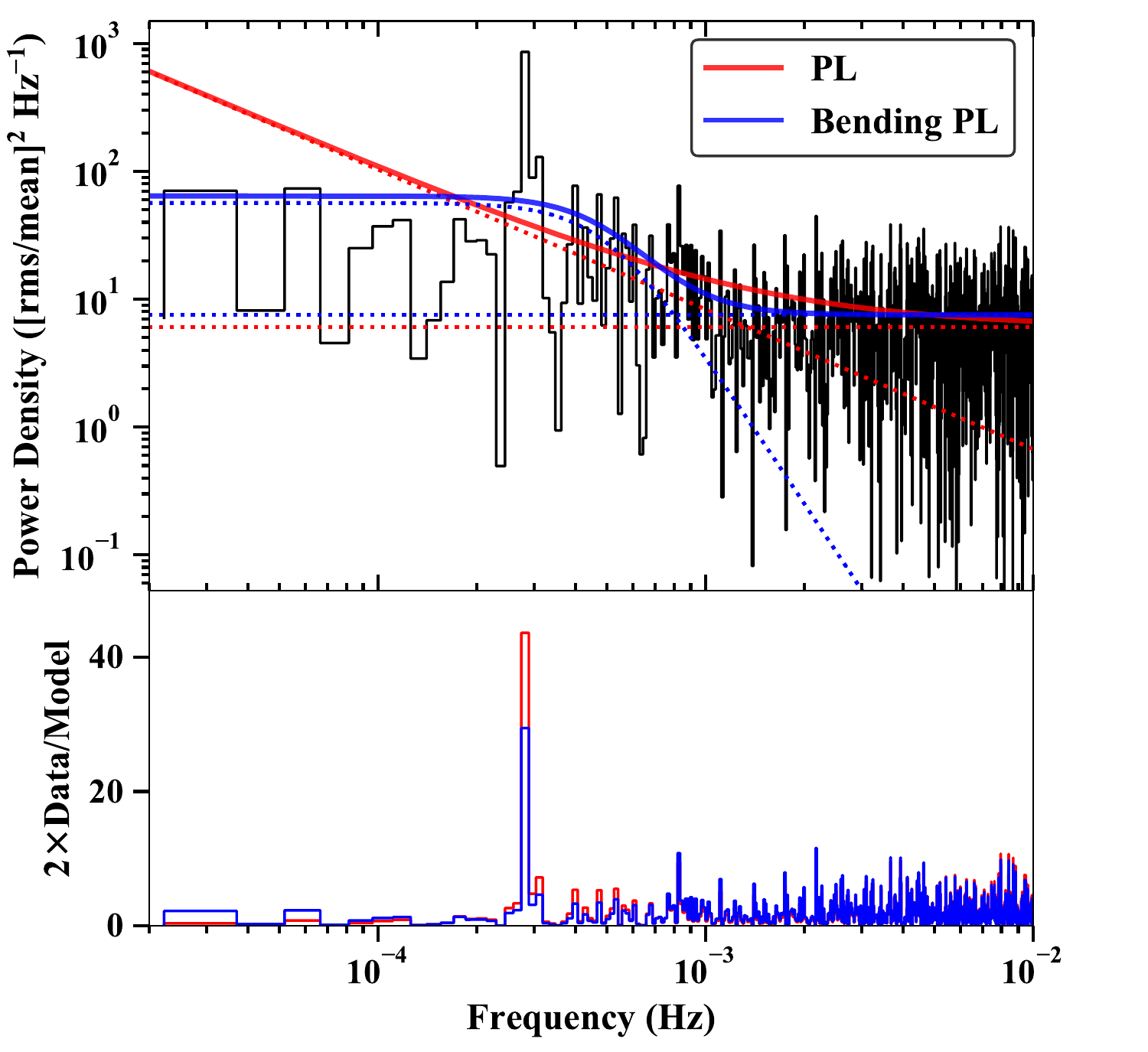}
\caption{The 1-4 keV PSD of \rej1034\ in Obs-9, fitted with a single power law model (RL, red), or a bending power law model (Bending PL, blue). The high frequency range is dominated by the Poisson noise power, which is modeled as a free constant. The solid and dash lines indicate the total models and their separate components. The lower panel shows the data-to-model ratio (times by 2) vs. frequency, where the QPO feature is clearly visible in both models.}
\label{fig-nulltest}
\end{figure}

\begin{figure*}
\centering
\includegraphics[trim=0.0in 0.3in 0.0in 0.0in, clip=1, scale=0.57]{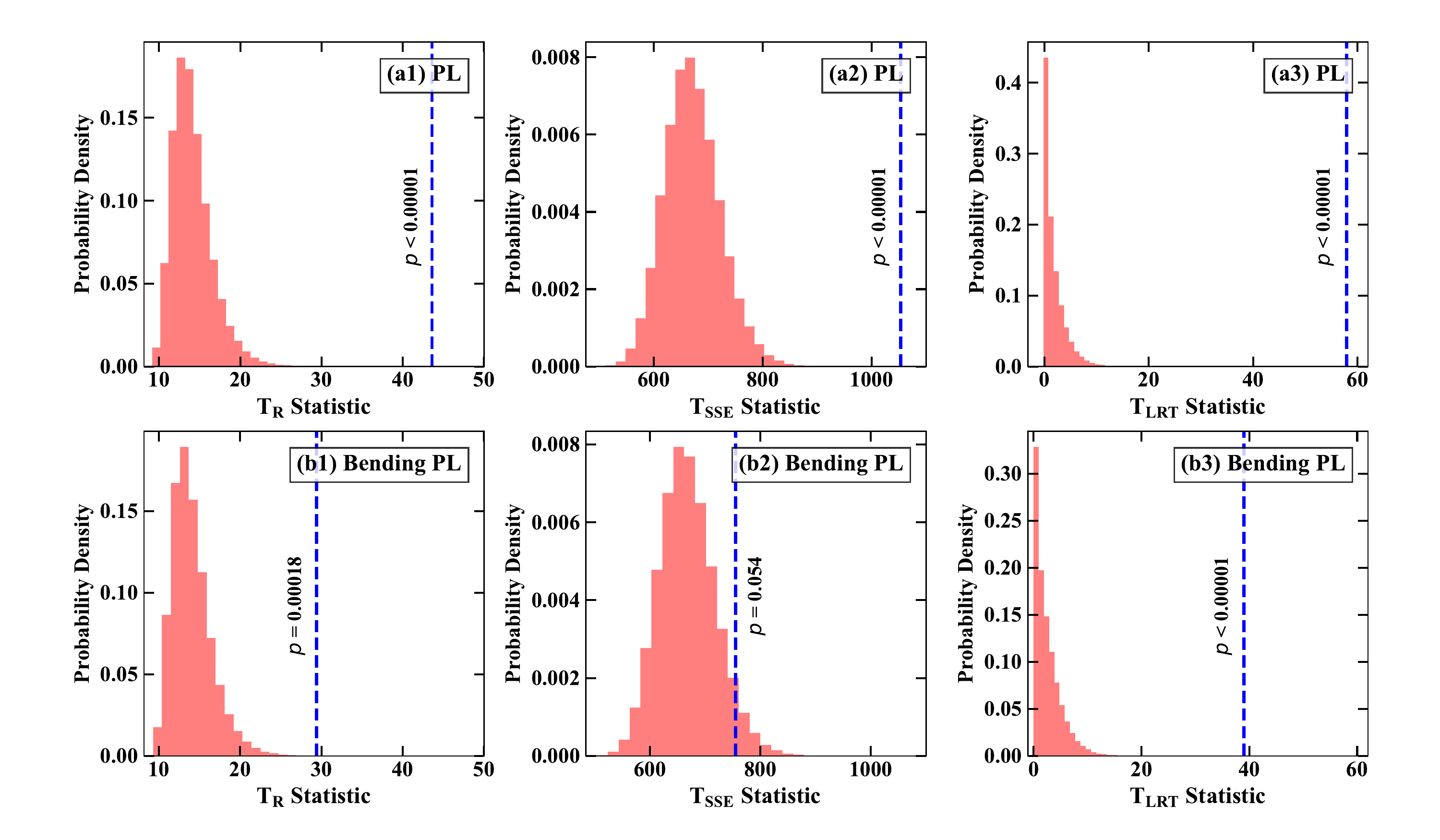}
\caption{{\it posterior} predictive distributions of the $T_{\rm R}$ and $T_{\rm SSE}$ statistics for the power law model (PL) and the bending power law model (Bending PL) for the 1-4 keV PSD of \rej1034\ in Obs-9. The $T_{\rm LRT}$ statistic is for checking the decrease of deviance after adding a lorentzian component to the continuum model to fit the QPO feature. The observed value is shown by the vertical blue dash line, together with the corresponding {\it posterior} predictive $p$-value. These results suggest that the QPO feature seen in Figure~\ref{fig-nulltest} should be an intrinsic component of the PSD.}
\label{fig-nullhist}
\end{figure*}

\begin{table}
\centering
   \caption{The first row shows the $R_{\rm QPO}$ value (i.e. 2$\times$data/continuum at the QPO frequency) measured in the PSD of \rej1034\ in Obs-9. As shown in Figure~\ref{fig-psd}, we use a power law plus a Poisson noise constant and a lorentzian profile to model the entire PSD. The critical $R_{\rm QPO}$ values for different confidence limits are derived from our Bayesian PSD simulations. The final row shows the significance of the observed QPO.}
\begin{tabular}{lcccc}
\hline
	 & 0.3-1 keV & 1-4 keV & 2-10 keV \\
\hline
$R_{\rm QPO, obs}$   & 39.8 & 93.3 & 22.2 \\
\hline
$R_{\rm QPO, 2\sigma}$	& 6.4 & 6.4 & 7.1 \\
$R_{\rm QPO, 3\sigma}$	& 12.6 & 12.6 & 14.1 \\
$R_{\rm QPO, 4\sigma}$	& 21.4 & 21.1 & 24.1 \\
Sig. of $R_{\rm QPO, obs}$ & $5.7\sigma$ & $9.0\sigma$ & $3.8\sigma$ \\
\hline
\end{tabular}
\label{tab-qpo-sig}
\end{table}

\begin{table}
\centering
   \caption{Results of the MLE fit and Bayesian analysis of the PSDs of \rej1034\ in different energy bands. $f_{\rm QPO}$ is the peak frequency of the best-fit lorentzian profile to the QPO. $W_{\rm QPO}$ is the FWHM of the best-fit QPO lorentzian profile in the log-log space. rms$_{\rm QPO}$ is the rms of the QPO by integrating the best-fit lorentzian profile. $\alpha_{\rm pl}$ is the slope of the continuum noise fitted by a power law. {\it Pos}  is the Poisson noise power. We also list values corresponding to the Bayesian mean, 5\% and 95\% percentiles.}
\begin{tabular}{lcccc}
\hline
Parameter & Method & 0.3-1 keV & 1-4 keV & 2-10 keV \\
\hline
$f_{\rm QPO}$ & MLE & 2.83 & 2.83 & 2.87 \\
($\times10^{-4}$ Hz)	& 1$\sigma$ & 0.06 & 0.07 & 0.08 \\
			& Mean & 2.80 & 2.82 & 2.91 \\
			& 5\%  & 2.63 & 2.67 & 1.26 \\
			& 95\% & 2.96 & 2.97 & 7.40 \\
\hline
$W_{\rm QPO}$ & MLE & 0.014 & 0.018 & 0.017 \\
Log (Hz)		& Mean & 0.008 & 0.012 & 0.013 \\
			& 5\% & 2.1E-7 & 2.0E-6 & 3.4E-9 \\
			& 95\% & 0.044 & 0.057 & 0.074 \\
\hline
rms$_{\rm QPO}$ & MLE & 4.0 & 12.4 & 10.8 \\
(\%)			& Mean & 4.1 & 12.3 & 11.0 \\
			& 5\%  & 1.5 & 6.7 & 4.7 \\
			& 95\% & 6.8 & 19.1 & 17.4 \\
\hline
$\alpha_{\rm pl}$ & MLE & -1.29 & -0.71 & -0.37 \\
			& Mean & -1.39 & -0.99 & -0.52 \\
			& 5\%  & -1.06 & -0.37 & -0.18 \\
			& 95\% & -1.71 & -1.72 & -0.96 \\
\hline
{\it Pos} & MLE & 0.63 & 6.05 & 12.95 \\
			& Mean & 0.67 & 3.93 & 3.74 \\
			& 5\%  & 0.55 & 2.90 & 2.7E-3 \\
			& 95\% & 1.37 & 17.9 & 24.8 \\
\hline
\end{tabular}
\label{tab-psd}
\end{table}

\begin{figure*}
\centering
\includegraphics[trim=0.05in 0.4in 0.0in 0.2in, clip=1, scale=0.5]{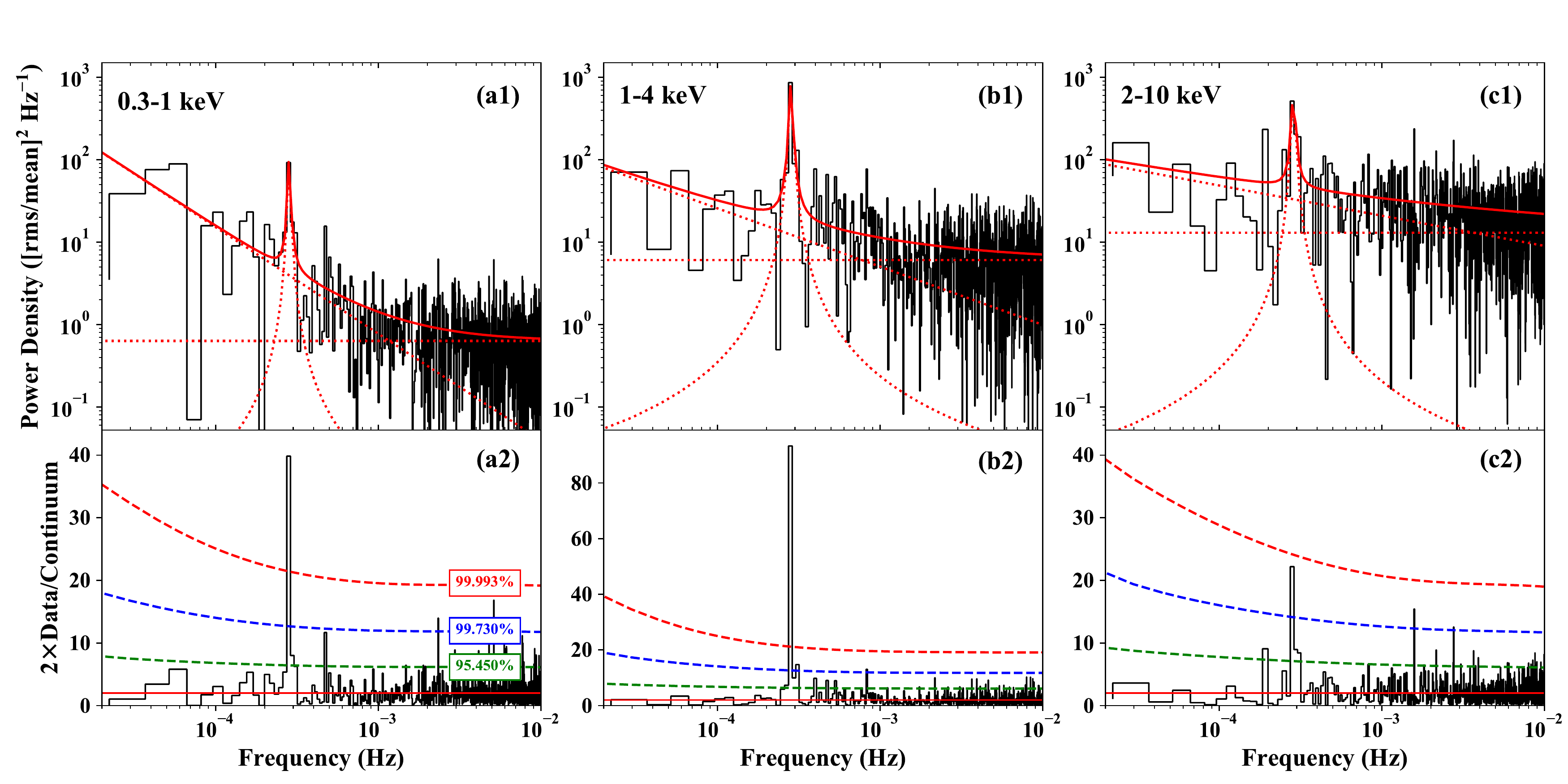} 
\caption{PSDs of \rej1034\ in the 0.3-1, 1-4 and 2-10 keV bands, respectively. In Panel a1, the red solid line indicates the best-fit model which is decomposed into three red dotted lines, including a power law to fit the intrinsic underlying noise, a free constant to fit the Poisson noise, and a lorentzian profile to fit the QPO signal. In Panel a2, the ratio of the PSD data to the best-fit PSD continuum is shown. A strong and coherent QPO peak can be seen at $\sim2.8\times10^{-4}$ Hz (see Table~\ref{tab-psd} for more accurate values). The green, blue and red dashed lines indicate the 2, 3 and 4 $\sigma$ confidence limits of the fluctuation in the red noise, respectively, with the model assuming that the QPO is a real PSD component superposed on the red noise continuum (see Section~\ref{sec-bayesian}).}
\label{fig-psd}
\end{figure*}

\subsection{X-ray PSD and the QPO Signal}
\label{sec-psd}
In order to quantitatively measure the QPO signal, we perform analysis in the frequency domain. We first produce the PSD{\footnote{This is actually a periodogram, which is a single realization of the intrinsic PSD. In this work we simply call it a PSD.}} for the 1-4 keV light curve, where the QPO appears more significant than in other bands (Alston et al. 2014). The first 4 ks data are excluded because of the severe background contamination. The normalization of these PSD is chosen such that the integration of the PSD is the fractional rms variability (i.e. the Belloni-Hasinger normalization, Belloni \& Hasinger 1990). Indeed, a strong peak feature can be identified in the PSD, as shown in Figure~\ref{fig-nulltest}. The frequency of this feature is similar to previously reported values (Alston et al. 2014), implying that it should be the same QPO signal as found in observations before 2011. The QPO feature is very narrow, although there appears to be a broader base which may be partly due to the fluctuation of the underlying red noise. For the width of the smallest frequency sampling interval, the quality factor of the QPO is 20 in the 1-4 keV band, suggesting that this QPO signal is highly coherent.

\subsubsection{Testing Continuum-only Hypothesis for the PSD}
We then perform some null hypothesis tests. Firstly, we fit a single power law model to the PSD using the maximum likelihood estimates (MLE) method (Vaughan 2010). Under the Belloni-Hasinger normalization, the theoretical Poisson power is a constant value, so we add a free constant to the power law and let the fitting determine the Poisson power. The red lines in the upper panel of Figure~\ref{fig-nulltest} shows the best-fit PSD model and the separate components.
The power law slope is found to be -1.09. The Poisson noise power dominates above $10^{-3}$ Hz, while the red noise power dominates lower frequencies. A standard way to estimate 
the significance of the observed power, $I_{\rm j}$, deviation from the model continuum, $S_{\rm j}$, at any frequency  $f_{\rm j}$ is $R_{\rm j}=2 I_{\rm j}/S_{\rm j}$. This can be used to make a test statistic 
$T_{\rm R}={\rm max}(R_{\rm j})$. This is shown in the lower panel of Figure~\ref{fig-nulltest}. The QPO is obvious, with the observed $T_{\rm R}$ being 43.6.

However, this does not simply give a significance of the QPO via the $\chi^2$ distribution with two degrees of freedom (dof) of the observed power $I_{\rm j}$, because there are also uncertainties in the model $S_{\rm j}$ which should be taken into account. Instead, we follow the more robust Bayesian prescription of Vaughan (2010) which includes the uncertainty of estimating the intrinsic PSD parameters in the simulated {\it posterior} predictive periodograms. 

We simulate the continuum model using the initial values of the MLE parameters, assuming a uniform prior probability density function (Vaughan 2010; Alston et al. 2014). The Python {\sc emcee} package is used to perform the Markov Chain Monte Carlo (MCMC) sampling in order to draw from the {\it posterior} of model parameters (Hogg, Bovy \& Lang 2010). We generate $10^{5}$ {\it posterior} predictive periodograms, and fit each of them with the same model. Then the {\it posterior} predictive distributions (PPDs) are derived for $T_{\rm R}$. These are shown in Figure~\ref{fig-nullhist} Panels-a1. The {\it posterior} predictive {\it p}-value of $T_{\rm R}$ is $<~10^{-5}$, i.e. none of the simulated periodograms can produce a larger $T_{\rm R}$ than the observation.

These same simulations also allow us to assess the overall goodness of fit of the power law continuum model to the data.
The fit has overall $\chi^2=1053.5$, so this sum of squared standard errors can be used as a test statistic $T_{\rm SSE}=\chi^2$. The PPDs of $T_{\rm SSE}$ are shown in Figure~\ref{fig-nullhist} Panels-a2. The {\it posterior} predictive {\it p}-value of $T_{\rm SSE}$ is $<~10^{-5}$, i.e. none of the simulated periodograms can produce a larger $T_{\rm SSE}$.
These results clearly rule out the power law continuum-only null hypothesis.

The X-ray PSD of AGN often shows a break at high frequencies (M$^{c}$Hardy et al. 2006, 2007). Vaughan (2010) shows that a bending power law is a better fit than a power law for the PSD of \rej1034, when the QPO feature is not modeled separately. Thus we replace the power law model with a bending power law{\footnote{A lower limit of 0 is put to the low-frequency slope of the bending power law model in order to maintain a realistic PSD shape of AGN.}}, and repeat all the above analysis. The MLE bend frequency is $4.9\times10^{-4}$ Hz. This slightly reduces $T_{\rm R}$ to 29.4 i.e. the QPO significance is still high, but has more impact on the overall fit quality, with the $T_{\rm SSE}$ being 812.4. The PPDs of $T_{\rm R}$ and $T_{\rm SSE}$ are shown in Figure~\ref{fig-nullhist} Panels-b1 and b2. The {\it posterior} predictive {\it p}-values for the two statistics are 0.00018 and 0.054. The bending power law model does give a better overall fit to the PSD which is within the 95\% confidence limit, but the deviation at the QPO frequency  is still significant at the 0.00018 level. 

Therefore, we can conclude that neither a power law nor a bending power law model can fully describe the PSD of \rej1034. The peak feature in $(2.5-3.5)\times10^{-4}$ Hz is clearly an intrinsic signal in the PSD. Therefore, it would not be appropriate to include this feature in the PSD's continuum fitting, and the previous suggestion regarding the bending power law being preferred over a power law is no longer valid. Indeed, if we mask out this band from the fitting, we find that there is no statistical difference between a power law and a bending power law. Below we show the results when this QPO-like feature is modelled independently.

\subsubsection{More Complete PSD Models}
We now add a Lorentzian component to the model to describe the QPO-like feature, and test for the significance of this additional component using a likelihood ratio test statistic, $T_{\rm LRT}$, which is derived from the difference in $\chi^2$ between the model with and without the Lorentzian. We emphasize that our application of $T_{\rm LRT}$ does not require the two models to be nested (Vaughan 2010). $T_{\rm LRT}$ is found to be very large for the power law continuum, with a value of 57.9. The previous MCMC simulations of models for the continuum were fit with both a continuum and a continuum plus Lorentzian component, and the PPD for the change in $\chi^2$ for a model including a Lorentzian is shown in Figure~\ref{fig-nullhist} Panel-a3 for the power law and Panel-b3 for the bending power law. Both have {\it posterior} predictive {\it p}-values of $T_{\rm LRT}$ $<~10^{-5}$, This shows that a PSD model with a separate QPO component is significantly better than a continuum-only model at the level of $<~10^{-5}$.

As a further test, we also compare the goodness of fit between the power law plus Lorentzian and the previous bending power law-only model. The observed $T_{\rm LRT}$ between these two models is 23.1, with the bending power law-only model being the less preferred model. Then we perform $10^5$ simulations of the {\it posterior} predictive periodograms using the bending power law-only model. For all the simulated periodograms, the bending power law-only model is always better than the power law plus Lorentzian model. Hence this is further evidence that the observed PSD must be very different from a single bending power law.

In addition to the above statistical tests, it is also important to emphasize that so far this QPO has been repeatedly detected at similar frequencies in 6 out of 9 independent \xmm\ observations (see Table~\ref{tab-obs}), hence it must surely be a real signal intrinsic to the source, rather than a temporary feature due to the fluctuation of the underlying red noise. 

In order to model these PSDs, we first need to decide which PSD continuum model to use. Previous works reported that for \rej1034\ a bending power law model fits the PSD better than a single power law model (Vaughan 2010; Alston et al. 2014). However, it is important to realize that the bending of the power law was mainly driven by the QPO feature which was never modeled as a separate component. But now the QPO is confirmed to be an intrinsic PSD component, we should include it in the model and test the continuum PSD model again. 

We compare the power law model with a bending power law under the condition that the QPO is additionally modeled by a separate Lorentzian. In this case, the $T_{\rm LRT}$ statistic between the two models is only 11.5, which corresponds to a {\it posterior} predictive {\it p}-value of 0.07. This indicates that the difference between the two model fits is not very significant. Also, the best fit bend frequency is found to be $9.7\times10^{-4}$ Hz. This frequency is two orders of magnitude higher than $1.7\times10^{-5}$ Hz, which is estimated from the correlation between the break frequency, black hole mass and optical luminosity of AGN (M$^{c}$Hardy et al. 2006), for a black hole mass of $2\times10^{6}~M_{\odot}$ and optical luminosity of $2\times10^{43}$ erg s$^{-1}$ (Jin et al. 2012a). A similar low break frequency at $\sim~10^{-5}$ Hz is also seen in Chaudhury et al. (2018) in their longer timescale broadband PSD of \rej1034. Thus it is unlikely that this best-fit bend is an intrinsic feature of the PSD,
especially as it is not present in any of the other energy bands (see Figure~\ref{fig-psd}a and c) or previous observations (Alston et al. 2014).
Hence, we adopt a PSD model of {\sc powerlaw + Lorenzian + constant} for all the subsequent fits.

\begin{figure*}
\centering
\begin{tabular}{cc}
\includegraphics[trim=-0.1in 0.2in 0.0in 0.0in, clip=1, scale=0.55]{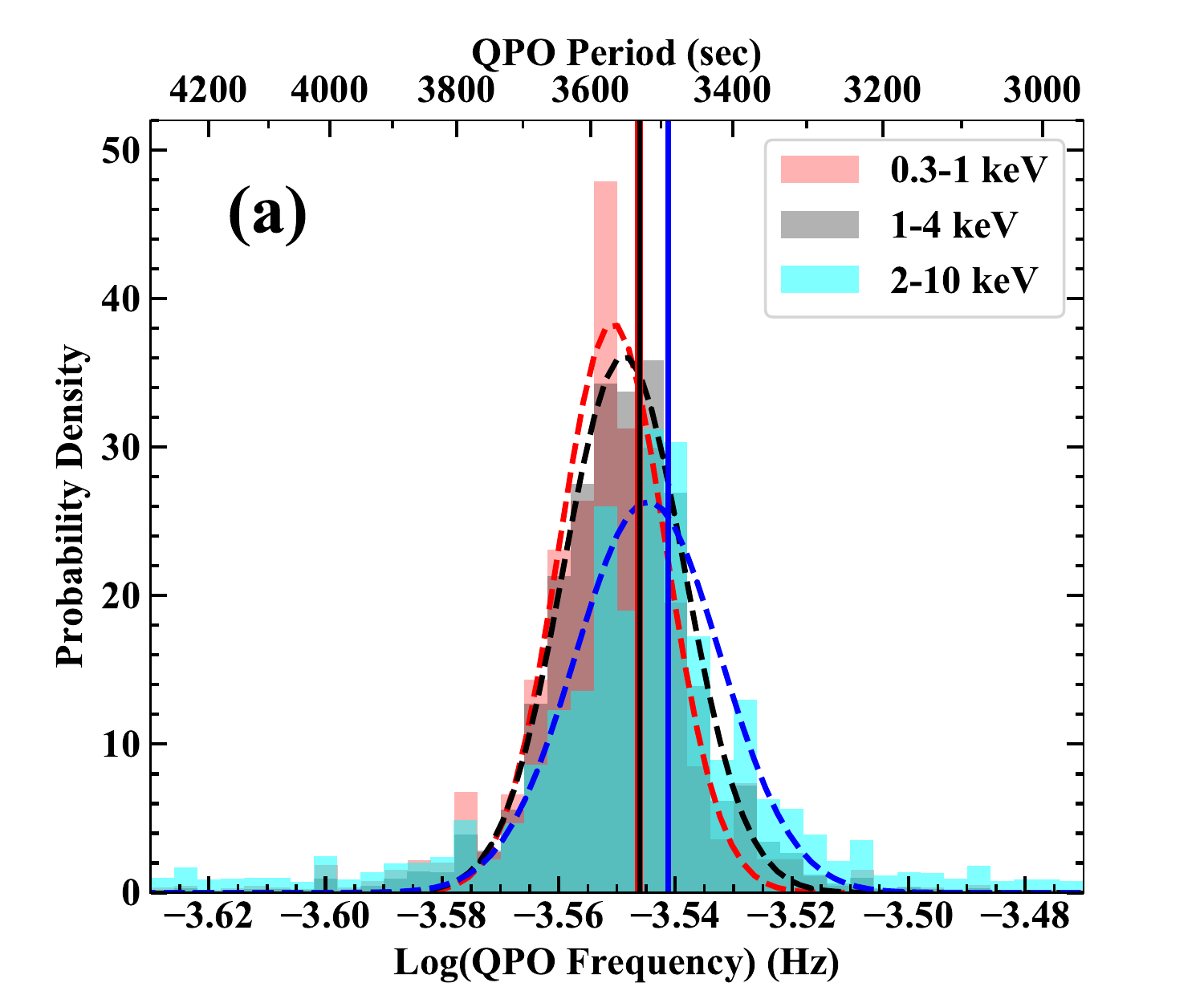} &
\includegraphics[trim=-0.1in 0.2in 0.0in 0.0in, clip=1, scale=0.55]{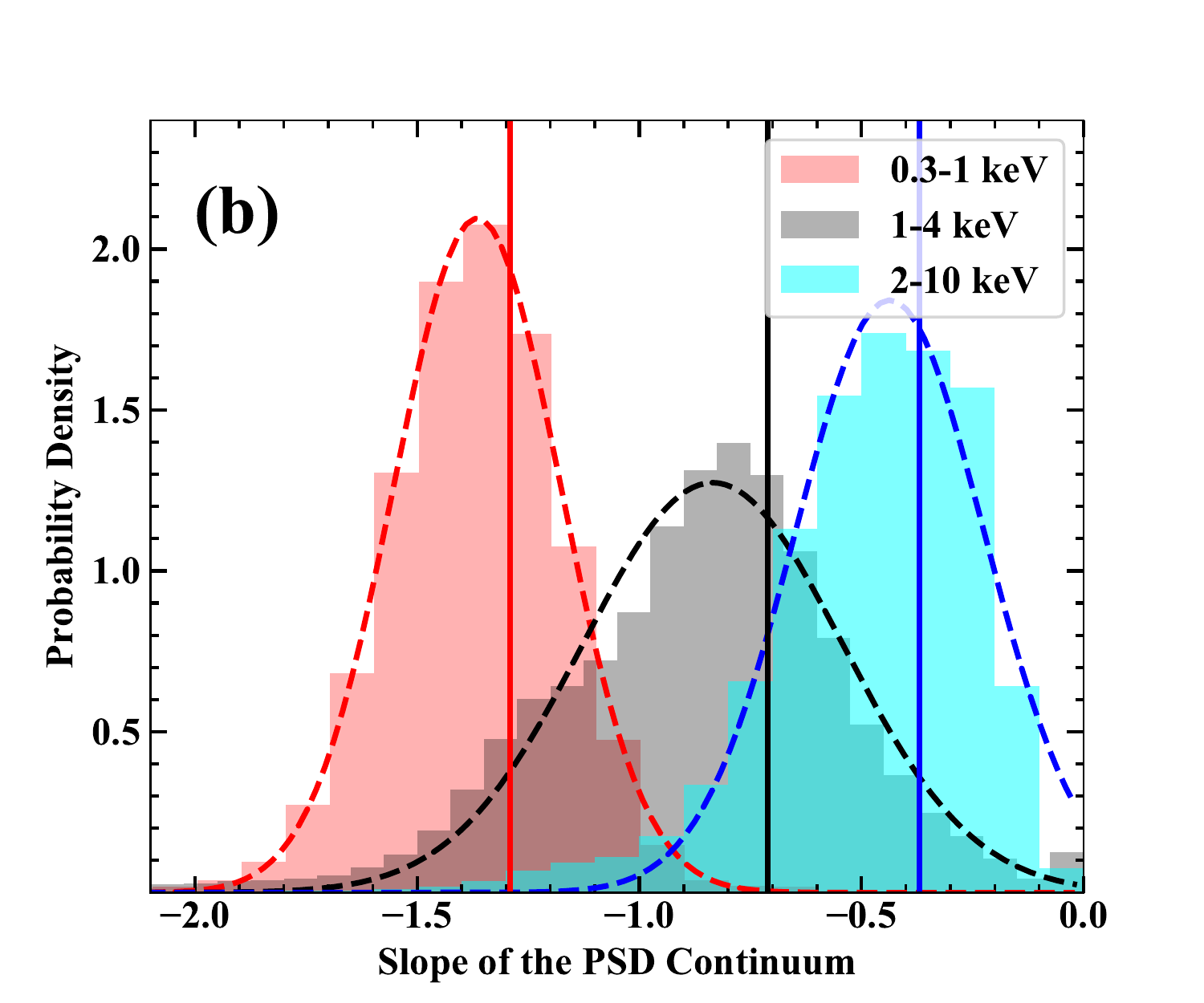} \\
\end{tabular}
\caption{The {\it posterior} predictive distributions of the QPO frequency (Panel-a) and the slope of the PSD continuum (Panel-b) in 0.3-1 keV (red histogram), 1-4 keV (black histogram) and 2-10 keV (cyan histogram). The best-fit Gaussian profiles of these distributions are shown by the dash lines. The vertical solid lines mark the best-fit MLE values (see Section~\ref{sec-bayesian} for detailed descriptions).}
\label{fig-qpomc-ene}
\end{figure*}

\section{Detailed PSD and QPO Analyses}
\subsection{Energy Dependence of the PSD and QPO}
\label{sec-bayesian}
In order to further explore the PSD and QPO, we 
examine their behaviours in different energy bands. Firstly, we produce PSDs for the light curves in 0.3-1, 1-4 and 2-10 keV bands. The upper panels of Figure~\ref{fig-psd} show that a QPO feature exists in the PSDs of all three energy bands at similar frequencies, and all of them appear very narrow. Then we fit these PSDs with the PSD model mentioned above. The MLE model fits are shown 
as the red lines in the upper panels of Figure~\ref{fig-psd}. The best-fit parameters are determined by the MLE method and are listed in Table~\ref{tab-psd}. The Poisson noise power is higher in harder X-rays because of the decreasing count rate.

The lower panels of Figure~\ref{fig-psd} show the ratio $R_{\rm j}=2\times I_{\rm j}/S_{\rm j}$, where $S_{\rm j}$ is the continuum model only, i.e. the same power law plus Poisson constant but {\em without} including the Lorentzian. The $R_{\rm j}$ value at the QPO frequency in the 1-4 keV band now increases to 93.3, higher than the value of 43.6 in the previous section due to the power law continuum level being lower when the Lorentzian is separately modelled.

We perform Bayesian analysis on the continuum power law (plus noise) with MCMC sampling as in Section~\ref{sec-obs9} to produce $10^5$ {\it posterior} predictive periodograms, and use these to put the $2,3$ and $4\sigma$ significance levels (dashed lines) on the lower panels of Figure~\ref{fig-psd}. Since we only have $10^5$ simulations, we cannot go beyond a probability of $10^{-5}$ i.e. $4.6\sigma$. However, we can assess the peak significance by scaling the 
PPD results e.g. for the 1-4~keV energy band, the $R$ values for $2, 3, 4\sigma$ 
are 6.4, 12.6 and 21.1. By comparison, a standard $\chi^2$ distribution with 2 dof has $R$ values corresponding to these $\sigma$ levels of 6.2, 11.8 and 19.3, which are only slightly smaller than for the full simulations. The peak 
$R$ is 93.3 in this band, which is $4.4\times$ larger than the $R$ value at 
$4\sigma$. For a standard
$\chi^2$ distribution with 2 dof, a $R$ value which is $4.4\times$ higher than that for $4\sigma$ would be a $9\sigma$ significance.
We similarly assess the significance level of the QPO in the 0.3-1~keV band to be $5.7\sigma$, but for the 2-10~keV band the PPD directly give the significance as $3.8\sigma$. The $R$ values and significances for each energy band are listed in Table~\ref{tab-qpo-sig}.
We emphasize that the confidence limits for the $R$ value in Figure~\ref{fig-psd} are used to assess the significance of the QPO at any particular frequency. Thus it is different from the $T_{\rm R}$ test presented in the previous section, which is used to assess the significance of a QPO signal over the entire frequency band.

Note that before this study, the highest $R_{\rm QPO}$ for \rej1034\ was reported to be $\sim$60 for the 0.3-10 keV band in Obs-2 (Gierli\'{n}ski et al. 2008, using the same model of a power law continuum plus Lorentzian and Poisson noise). Therefore, not only does our new observation demonstrate the long-term nature of the QPO, but it also finds the so-far highest level of significance for an X-ray QPO signal in all AGN. 

We repeat the Baysian analysis with MCMC sampling on the full PSD model (including the Lorentzian). We use these PPD to set the 5\% and 95\% uncertainty ranges on the MLE parameter values for the power spectral components, as detailed in Table~\ref{tab-psd}. We show the full PPD for the QPO frequency in each energy band in Fig~\ref{fig-qpomc-ene}a. Clearly this is consistent across all energies, which is also confirmed by the overlap of the QPO frequency uncertainty ranges in Table~\ref{tab-psd}.

Table~\ref{tab-psd} shows that the 
width of the QPO is very small ($\Delta \log f = 0.014$ in 0.3-1 keV), and is 
consistent with being the same across all energy bands. The table also shows that
the fractional rms amplitude of the QPO 
increases significantly with energy, 
showing that a larger fraction of hard X-rays is varying at the QPO frequency than the soft X-rays. However, since the flux ratio between 0.3-1 keV and 1-4 keV is 7.5, the absolute rms amplitude of the QPO in 0.3-1 keV is actually larger than that in 1-4 keV by a factor of 2.4. The spectrum of the QPO will be examined in more detail in our next paper (Paper-II).

Fig~\ref{fig-qpomc-ene}b shows that the
best-fit power law slopes systematically harden at  higher energies, with $\alpha_{pl}=-1.29$ for 0.3-1 keV, -0.71 for 1-4 keV and -0.37 for 2-10 keV. Only 0.3\% of the simulations in 0.3-1 keV have power spectra as hard as observed in 1-4 keV, and only 0.01\% simulations in 0.3-1~keV have power spectra as hard as those observed in 2-10 keV. These results confirm that the steepening of the PSD continual slope towards softer X-rays is an intrinsic property of \rej1034.
However, the normalization of the power at low frequencies ($\sim 10^{-5}$~Hz) is $\sim 100$~[rms/mean]$^2$~Hz$^{-1}$, similar at all energies (see Figure~\ref{fig-psd}), which suggests that the difference is in the amount of high frequency power.

Similar properties of the PSD continuum are also seen in other NLS1s (e.g. Jin et al. 2013; Jin, Done \& Ward 2016, 2017a). This can be interpreted in a model where fluctuations propagate from the disc (which dominates at low energies) to the corona (which dominates at high energies), with additional fluctuations in the corona enhancing the high frequency power in the energy bands dominated by this component (e.g. Gardner \& Done 2014).

\subsection{Testing Potential Harmonics of the QPO Signal}
We also search for possible harmonics associated with this highly significant QPO. In BHBs, a high-frequency QPO may have harmonics at frequency ratios of 2:3, 3:5 and 2:5, and a low-frequency QPO may exhibit a harmonic frequency ratio of 1:2. Since it is not clear if the detected QPO in \rej1034\ represents the fundamental or harmonic frequency, we check all possible harmonic frequencies for potential peak features. Based on the observed QPO frequency of $2.8\times10^{-4}$ Hz, the 2:3, 3:5 and 2:5 ratios predicts potential peaks at $1.9\times10^{-4}$, $4.2\times10^{-4}$, $1.7\times10^{-4}$, $4.7\times10^{-4}$, $1.1\times10^{-4}$ and $7.0\times10^{-4}$ Hz. The 1:2 ratio predicts potential peaks at $1.4\times10^{-4}$ and $5.6\times10^{-4}$ Hz.

However, it is already clear from the lower panel of Fig~\ref{fig-psd} that there are no features above $3\sigma$ significance at any other frequency in any of the energy bands. The 
strongest feature which is even close to any of the potential harmonics listed above is 
at $1.9\times10^{-4}$ Hz in the PSD of 2-10 keV band (see Figure~\ref{fig-psd} Panel-c), but
this is only seen at $\sim~2.1~\sigma$. 
No other energy bands show peaks with $>~2\sigma$ significance at any of the potential harmonic frequencies. The feature at $4.8\times 10^{-4}$ Hz in the 0.3-1~keV band 
has $\sim~2.4~\sigma$ significance, but this frequency is not harmonically related. No significant harmonics are seen in the Obs-2 data
either (Gierli\'{n}ski et al. 2008; Vaughan et al. 2010; Alston et al. 2014). Therefore, we conclude that there are no significant harmonics of the QPO signal in the current data of \rej1034.

The QPO of \rej1034\ is often compared to the 67 Hz high-frequency QPO of the BHB \grs1915, 
as it approximately scales with the mass difference between these two accreting black holes (Middleton et al 2009). The overall energy spectra of \grs1915\ in observations showing the 67~Hz are very similar to those of \rej1034\ with a strong disc, a smaller warm Compton component, and an even smaller hot Compton tail (Middleton \& Done 2010). \grs1915\ shows
three harmonic peaks at 27 Hz (Belloni et al. 2001), 34 Hz (Belloni \& Altamirano 2013) and 40 Hz (Strohmayer 2001), but only the 
34 and 41 Hz  QPOs appear simultaneously with the 67 Hz QPO. 
The 34 Hz QPO has a fractional rms of 0.8\% and a quality factor of 13.1 in 2-15 keV. In comparison, the 67 Hz QPO has a rms of 2.0\% and a quality factor of 24.7 in the same energy band, and so the 34 Hz QPO is 60\% weaker than the 67 Hz QPO, but with a similar line width. The 41 Hz QPO has a fractional rms of 2.4\% and a quality factor of 7.7 in 13-27 keV, while in the same energy band the 67 Hz QPO has a rms of 1.9\% and a quality factor of 19.6, thus the 41 Hz QPO is 26\% more powerful than the 67 Hz QPO, but its profile is 56\% broader.

Assuming that the QPO of \rej1034\ has similar harmonics as the 67 Hz QPO in \grs1915, we can estimate that the intrinsic PSD of \rej1034\ may have an extra peak at $1.4\times10^{-4}$ Hz with a rms of 5.0\%, or at $1.7\times10^{-4}$ Hz with a rms of 15.6\%. Such features are not observed in the PSDs of \rej1034\ in Figure~\ref{fig-psd}. 
We test this explicitly using the 1-4 keV PSD. We add the expected harmonic at $1.4\times 10^{-4}$~Hz to the best fit PDS model and simulate $10^{5}$ periodograms. Only a fraction 0.052 of the simulations with the harmonic have power at that frequency as low as observed. We repeat the simulations for the potential harmonic at $1.7\times10^{-4}$ Hz, and find only a fraction 0.01 have power this small.
Therefore, the non-detection of these two potential harmonics is probably not due to the random fluctuation of the PSD swamping the signal, but rather it is intrinsic to \rej1034. 
The above analysis rules out the presence of harmonics of similar relative strengths to those observed in GRS \grs1915\ in the current observation of \rej1034, but we cannot rule out the possibility that much weaker harmonics may exist, but are swamped by the PSD's fluctuation. Furthermore, \grs1915\ does not often exhibit the 67 Hz QPO and its harmonics simultaneously. Clearly we cannot exclude the possibility that future observations of \rej1034\ may show these harmonics.

\section{Long-term Variation of the QPO}
We compare some key properties of the QPO between Obs-9 and previous observations, especially Obs-2 where the background contamination is low and the QPO signal can be detected across the entire 0.3-10 keV band. Such a comparison allows us to verify the robustness of various QPO properties, as well as checking if there is any evidence for the long-term variation of the QPO.

\subsection{Long-term Variation of the QPO Frequency}
\label{sec-qpovar-freq}
The QPO frequencies reported in previous observations are all in the range of $(2.5-2.7)\times10^{-4}$ Hz (see Table~\ref{tab-obs}), except for the new Obs-9, in which the frequency increases to $2.83\times10^{-4}$ Hz. Therefore, it is necessary to assess the significance of this difference. We only compare Obs-9 with Obs-2, because both of these two observations have low background, and the QPO signal is well determined. For Obs-2, Gerli\'{n}ski et al. (2008) reported that within the 23-83 ks data segment the QPO signal was more significant, thus we perform the comparison with the 0-83 ks and 23-83 ks data segments, separately. The data within 83-91 ks of Obs-2 are excluded because of the background contamination. Since the QPO frequency does not change significantly with the photon energy, we use the entire 0.3-10 keV data to achieve the best S/N in the light curve. The same Bayesian analysis is performed to derive the PPD of the QPO frequency.

\begin{figure*}
\centering
\begin{tabular}{cc}
\includegraphics[trim=0.0in 0.2in 0.0in 0.0in, clip=1, scale=0.56]{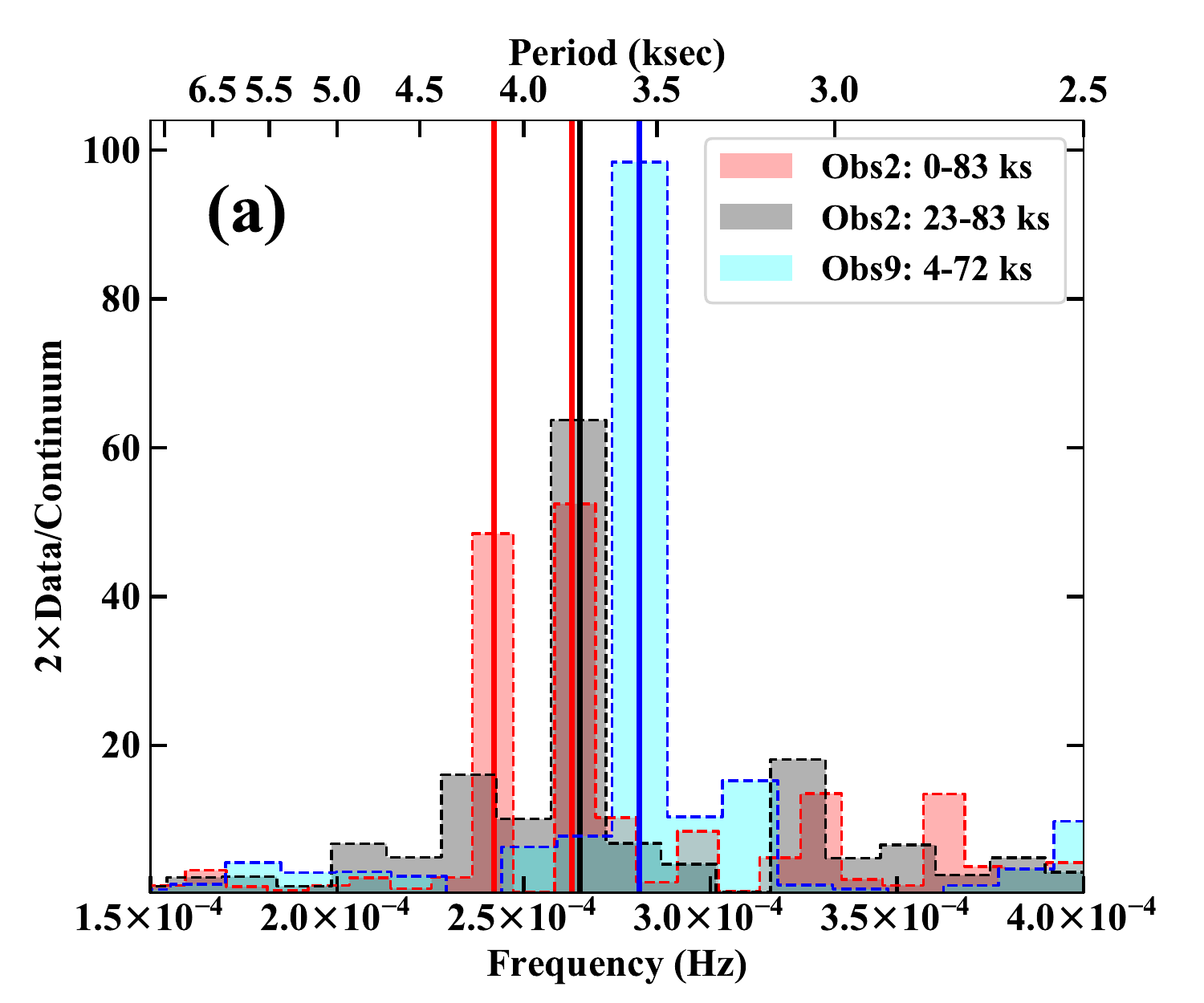} &
\includegraphics[trim=0.0in 0.2in 0.0in 0.0in, clip=1, scale=0.56]{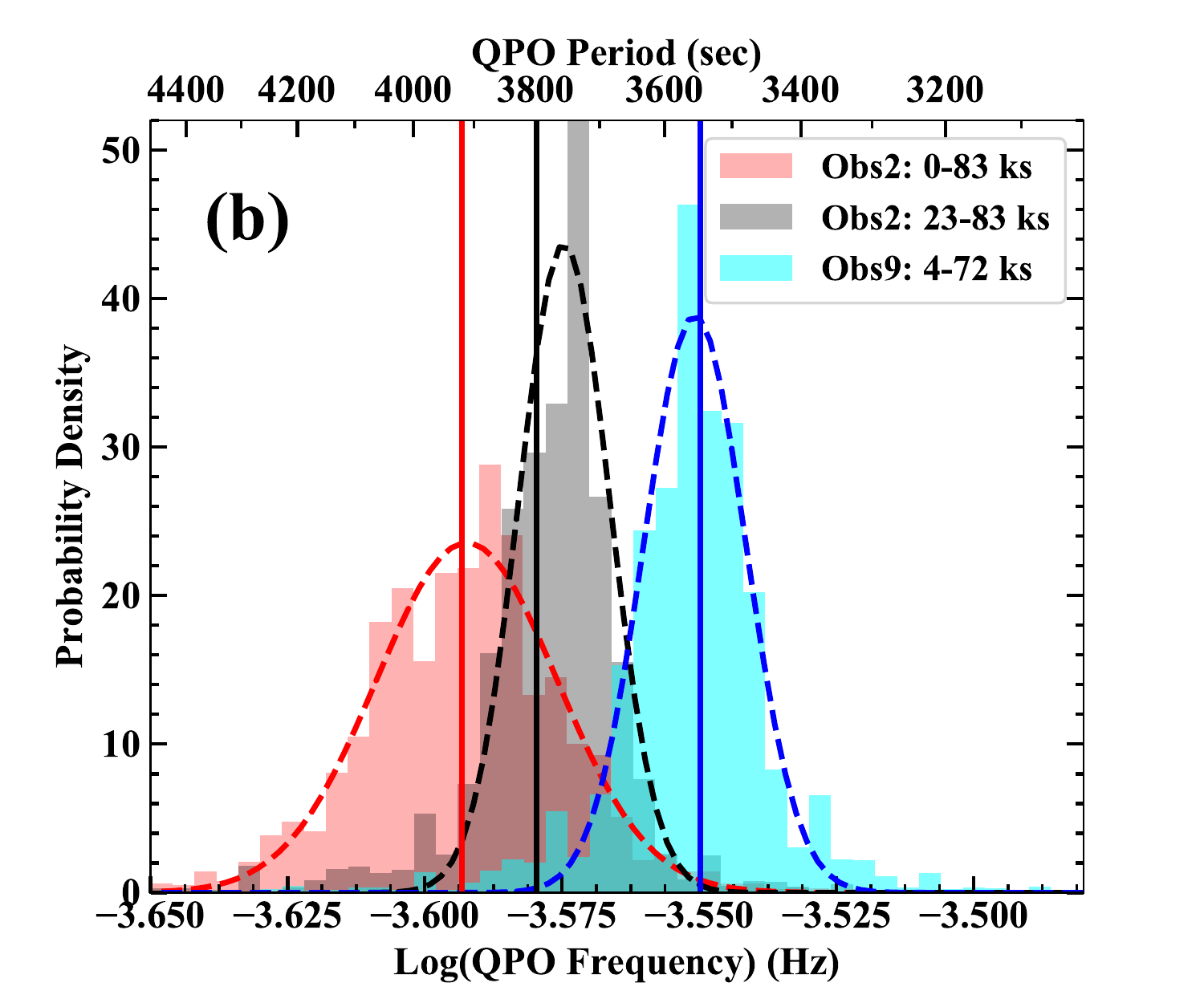} \\
\end{tabular}
\caption{Comparison of the QPO frequency observed in Obs-2 and Obs-9 for the 0.3-10 keV band. Panel-a: PSDs of different data segments relative to their best-fit continuum model around the QPO frequency. Solid vertical lines indicate frequencies of different QPO peaks. Two nearby narrow QPO peaks are seen in the 0-83 ks data segment of Obs-2. Panel-b: different histograms indicate the  {\it posterior} predictive distributions of the QPO frequency for different data segments. Dashed lines indicate the best-fit Gaussian profiles. Vertical solid lines indicate the best-fit MLE values.}
\label{fig-qpofreq-compare}
\end{figure*}

Figure~\ref{fig-qpofreq-compare} compares the QPO frequency between Obs-2 and Obs-9. In Panel-a, we compare the data-to-model ratio around the QPO frequencies for the three datasets. In Obs-2, when the 0-83 ks data segment is used, two nearby QPO peaks can be detected. The stronger peak is at $2.63\times10^{-4}$ Hz, while the weaker one is at $2.42\times10^{-4}$ Hz. If the first 23 ks data are excluded, the lower-frequency peak becomes much weaker. Hence, we think the low-frequency peak is mainly due to the instantaneous variation of the QPO period (Czerny et al. 2010; Hu et al. 2014). In comparison, the QPO in Obs-9 is clearly a single peak, and is shifted to a higher frequency. The histograms in Panel-b indicate the PPDs of the QPO frequency for the three datasets. The vertical solid lines indicate the best-fit MLE values. For Obs-2 the best-fit MLE period of QPO is $3920\pm150$ s in 0-83 ks, and $3800\pm70$ s in 23-83 ks, which are consistent with the results reported before (e.g. Gerli\'{n}ski et al. 2008; Alston et al. 2014; Hu et al. 2014). The QPO period in Obs-9 is, however, found to be $3550\pm80$ s, which is $250\pm100$ s (i.e. $\sim$7\% of the QPO period) smaller than in Obs-2. The difference between the PPDs of the two observations is also obvious. Compared to the PPD of the QPO frequency for the 0-83 ks segment of Obs-2, the QPO frequency in Obs-9 has a {\it posterior} predictive {\it p}-value of 0.019. For the 23-83 ks segment of Obs-2, the {\it p}-value is 0.028.

Based on these results, we report that the QPO frequency in Obs-9 is higher than that found in Obs-2. It is relevant to mention that the QPO also has a flickering nature within a single observation, and that the instantaneous period varies between 3000-4000 s (Hu et al. 2014). However, this does not mean that the observed long-term variation of the QPO frequency is simply due to the short-term variation. In fact, both Obs-2 and Obs-9 contain more than 20 QPO cycles, and so our comparison of the QPO frequency is statistically meaningful. However, it is not known if the increase of QPO frequency (e.q. the decrease of QPO period) within the last 11 years is a monotonic trend or a fluctuation, because in the other observations the QPO signal was not well constrained due to poor data quality (Alston et al. 2014). Clearly, future \xmm\ observations of \rej1034\ can bring further evidence on the long-term variation of the QPO frequency, and hence help us to underlying physical mechanisms involved.

\subsection{Reversed QPO Time-lag between Obs-2 and Obs-9}
\label{sec-qpovar-lag}
Another important phenomenon related to the QPO is the phase lag (eq. time lag in the time domain). Gierli\'{n}ski et al. (2008) applied the light curve folding method and found $\sim$260 s lag between 0.3-0.4 keV and 2-10 keV (leading) in Obs-2. Middleton, Done \& Uttley (2011) used the same data and method, and found $\sim370$ s lag between 0.2-0.3 keV and 1-10 keV (leading). However, we notice that this method is sensitive to the accuracy of the folding period. A more robust method is to perform the phase lag analysis in the Fourier domain (e.g. Uttley et al. 2014), because it differentiates the variability into different frequency bins. Zoghbi \& Fabian (2011) applied this method to the Obs-2 data, and found a lag of $\sim$500 s between 0.4-0.6 keV and 1.5-2.0 keV (leading), with a coherence of $\sim$0.6 around the QPO frequency. They also showed that the lag spectrum does not change significantly with the inner radius of the annular source extraction region, thereby ruling out any significant influence from pile-up.

\begin{figure*}
\centering
\includegraphics[trim=0.3in 0.2in 0.0in 0.3in, clip=1, scale=0.65]{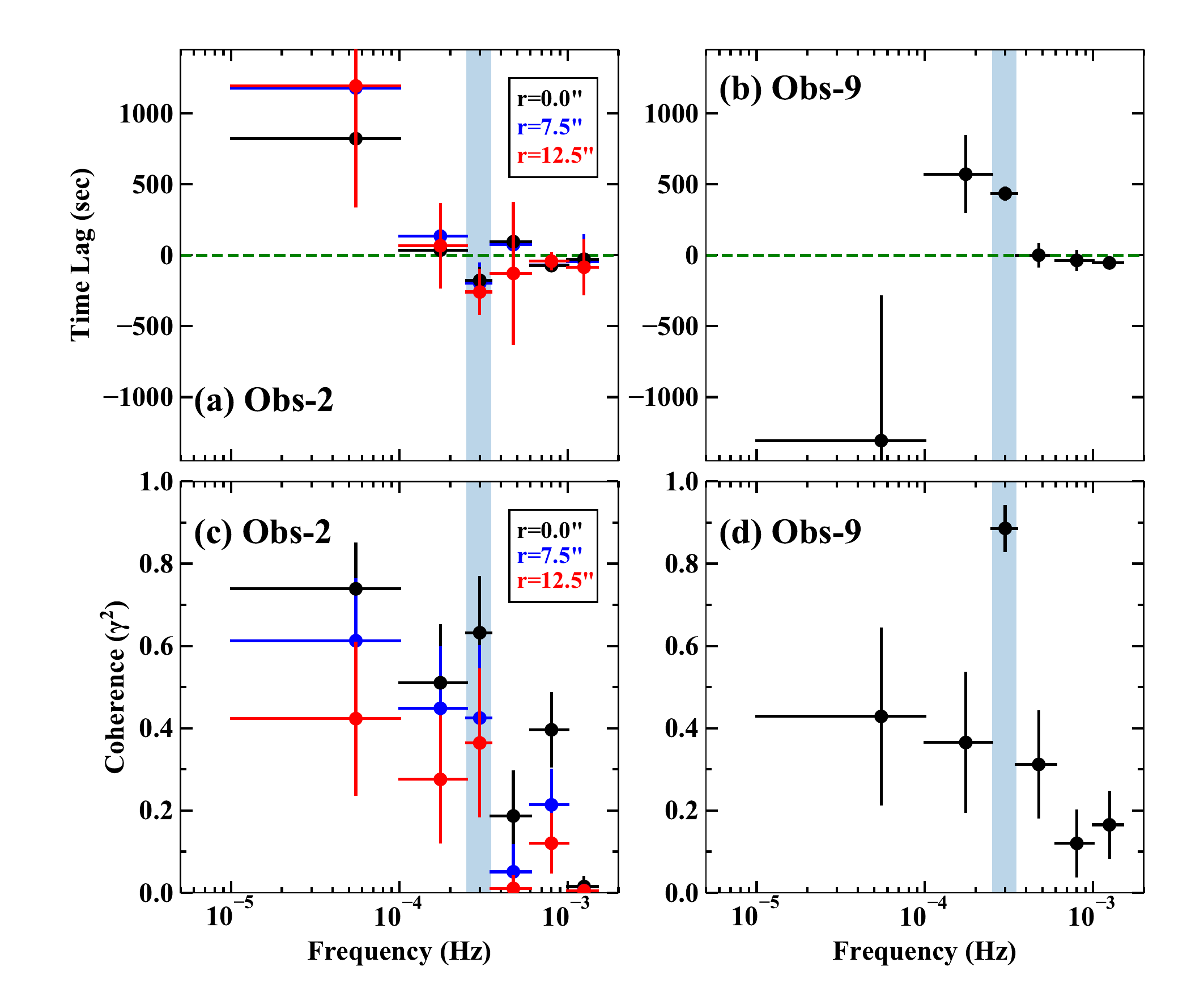}
\caption{Time-lag and coherence vs. frequency between the light curves in 0.3-1 keV and 1-4 keV for Obs-2 and Obs-9, separately. In each panel the shadowed region indicate the QPO frequency bin of $(2.5-3.5)\times10^{-4}$Hz. In Panels a and b, a positive lag means the soft X-ray variability leads the hard X-ray. In Panels a and c, the black, blue and red data points indicate the results for annular source extraction regions with inner radii being 0", 7.5" and 12.5", respectively.}
\label{fig-lag-compare}
\end{figure*}

For consistency, we first reproduce the above results for Obs-2. Three inner radii ($r_{\rm s}$) of the annular source extraction region are tested  in order to check the effect of pile-up. The resultant lag vs. frequency and coherence vs. frequency plots are shown in Figure~\ref{fig-lag-compare} Panels-a and c. The time lag and coherence values in the QPO frequency bin of $(2.5-3.5)\times10^{-4}$Hz are listed in Table~\ref{tab-qpo-lag}. Our analysis confirm that in Obs-2 the QPO in 0.3-1 keV lags behind 1-4 keV by 200-300 s for all values of $r_{\rm s}$ from  0" to 12.5". As the S/N drops towards larger inner radii, the lag becomes less significant with larger errors and the coherence becomes smaller (see Table~\ref{tab-qpo-lag}). However, even with $r_{\rm s}=$ 0" the lag is only detected at a significance of $2~\sigma$.

We then investigate the time lag in Obs-9, but without trying different source extraction regions as these data are not affected by pile-up. The results for lag vs. frequency and coherence vs. frequency are shown in Figure~\ref{fig-lag-compare} Panels-b and d. Surprisingly, we find that the lag in Obs-9 has an opposite direction from Obs-2, i.e. the QPO phase in the 1-4 keV band lags behind that in 0.3-1 keV band. The absolute lag value is $430\pm50$ s, which is much more significant than that found in Obs-2, and is also associated with a high coherence of $0.89\pm0.06$. Hence, the soft X-ray lead in Obs-9 is clearly a more significant and robust measurement than the soft X-ray lag found in Obs-2.

\begin{figure*}
\centering
\begin{tabular}{cc}
\includegraphics[trim=0.0in 0.0in 0.0in 0.0in, clip=1, scale=0.53]{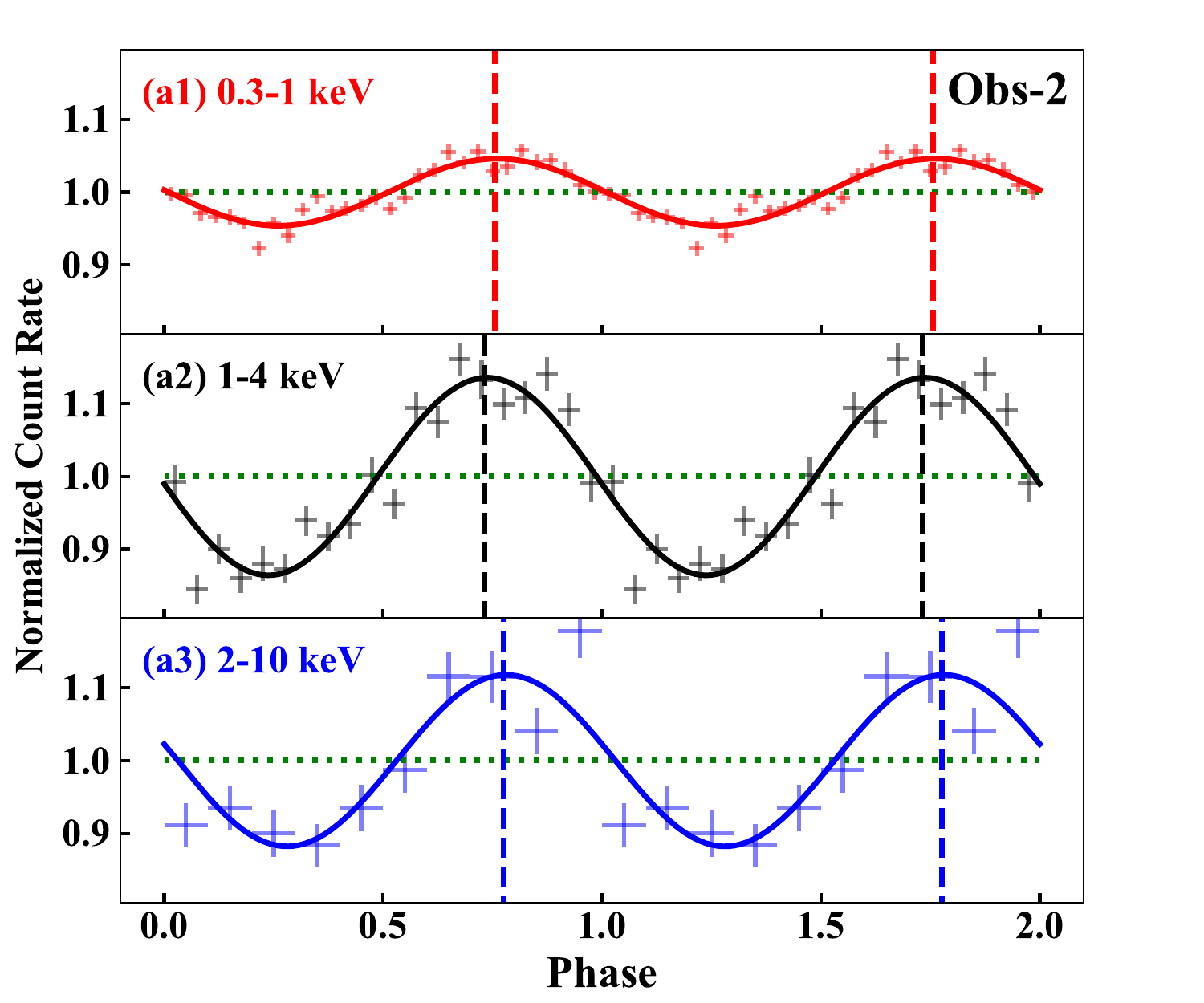} & 
\includegraphics[trim=0.0in 0.0in 0.0in 0.0in, clip=1, scale=0.53]{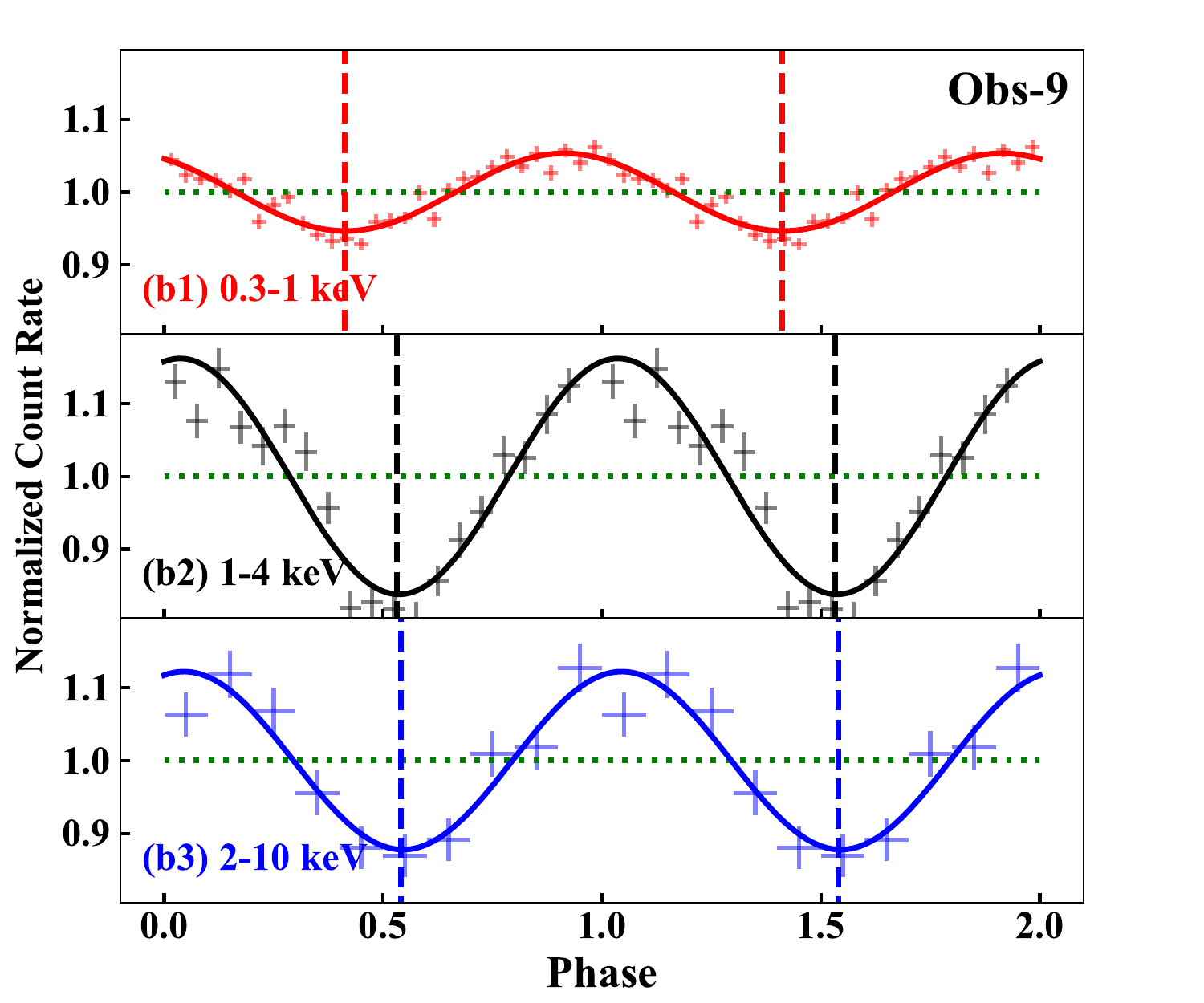} \\
\end{tabular}
\caption{Folded QPO light curves of \rej1034. Panels a1 to a3: folded QPO light curves in Obs-2 with a folding period of 3800 s. Two periods are shown to reveal the periodicity. The best-fit sinusoidal function with a phase shift are shown by the solid line in each panel. The vertical dash lines indicate the QPO peaks, where phase differences can be found between different energy bands. Panels b1 to b3: folded light curves in Obs-9 with a folding period of 3550 s.}
\label{fig-flc-compare}
\end{figure*}

As an additional check we apply the light curve folding method to the QPO in Obs-2 ($r_{\rm s}$=0") and Obs-9. We take the QPO period measured from the entire 0.3-10 keV band as the folding period, which is 3800 s for Obs-2 and 3550 s for Obs-9 (see Section~\ref{sec-qpovar-freq}). The folded light curves are produced for the 0.3-1, 1-4 and 2-10 keV bands, as shown in Figure~\ref{fig-flc-compare}. Indeed, we also find that the QPO phase in 0.3-1 keV lags behind the 1-4 keV band by 0.024 in Obs-2, while it leads the 1-4 keV band by 0.120 in Obs-9. Interestingly, we find that the QPO phase in 2-10 keV in Obs-2 also leads that for 1-4 keV by 0.044, although the S/N of the 2-10 keV band light curve is much lower. These results are consistent with the QPO lag analysis in the frequency domain.

The opposite time lag between Obs-2 and Obs-9 is difficult to understand. One possible explanation is the influence of the stochastic variability. Figure~\ref{fig-lag-compare} Panels-c and d show that in Obs-2 the coherence in the QPO frequency bin is smaller than the coherence at lower frequencies where the red noise dominates. This leads to the idea that the low-frequency stochastic variability suppresses the coherence in the QPO frequency bin, overwhelms the intrinsic lag of the QPO, and causes the apparent soft lag with a relatively low coherence. In comparison, the low-frequency stochastic variability in Obs-9 is weaker, and so its QPO shows a significant hard lag with a coherence that is much higher than all other frequencies. This red noise contamination can happen if there is a significant aliasing effect of the low-frequency power (Uttley et al. 2014). Additionally, the stochastic variability may also have a physical impact on the QPO properties (Czerny et al. 2010; Hu et al. 2014). Another possibility is the contamination of pile-up. Although it has been shown by Zoghbi \& Fabian (2011) and our analysis that the shape of the lag vs. frequency spectrum does not change significantly as the S/N decreases, the coherence does decrease significantly as more photons from the centre of the PSF are excluded, and a low coherence often means that the corresponding lag is not reliable. However, it is also possible that the phase lags observed in Obs-2 and Obs-9 are both real, in which case the reversed time lag would be an interesting new phenomenon. In any case, it is crucial to carry out further observations in order to understand the true lag behaviours of this QPO.

\begin{table}
\centering
\caption{The coherence and time lag between 0.3-1 keV and 1-4 keV in the QPO frequency bin of $(2.5-3.5)\times10^{-4}$ Hz for Obs-2 and Obs-9. $r_{\rm s}$ indicates different inner radii of the annular source extraction region for checking the pile-up effect. A positive lag indicates that the soft X-ray variability leads the hard X-ray. $N_{\rm data}$ indicates the number of data points in the periodogram being included in the QPO frequency bin.}
\begin{tabular}{llccc}
\hline
Obs & $r_{\rm s}$ & $N_{\rm data}$ & Time Lag & Coherence \\
\hline
Obs-2 & 0" & 9 & -180 $\pm$ 90 s & 0.63 $\pm$ 0.14 \\
Obs-2 & 7.5 & 9 & -200 $\pm$ 140 s& 0.42 $\pm$ 0.18 \\
Obs-2 & 12.5" & 9 & -260 $\pm$ 160 s& 0.36 $\pm$ 0.18 \\
Obs-9 & 0" & 7 & 430 $\pm$ 50 s & 0.89 $\pm$ 0.06 \\
\hline
\end{tabular}
\label{tab-qpo-lag}
\end{table}

\begin{figure}
\centering
\includegraphics[trim=0.1in 0.3in 0.0in 0.0in, clip=1, scale=0.5]{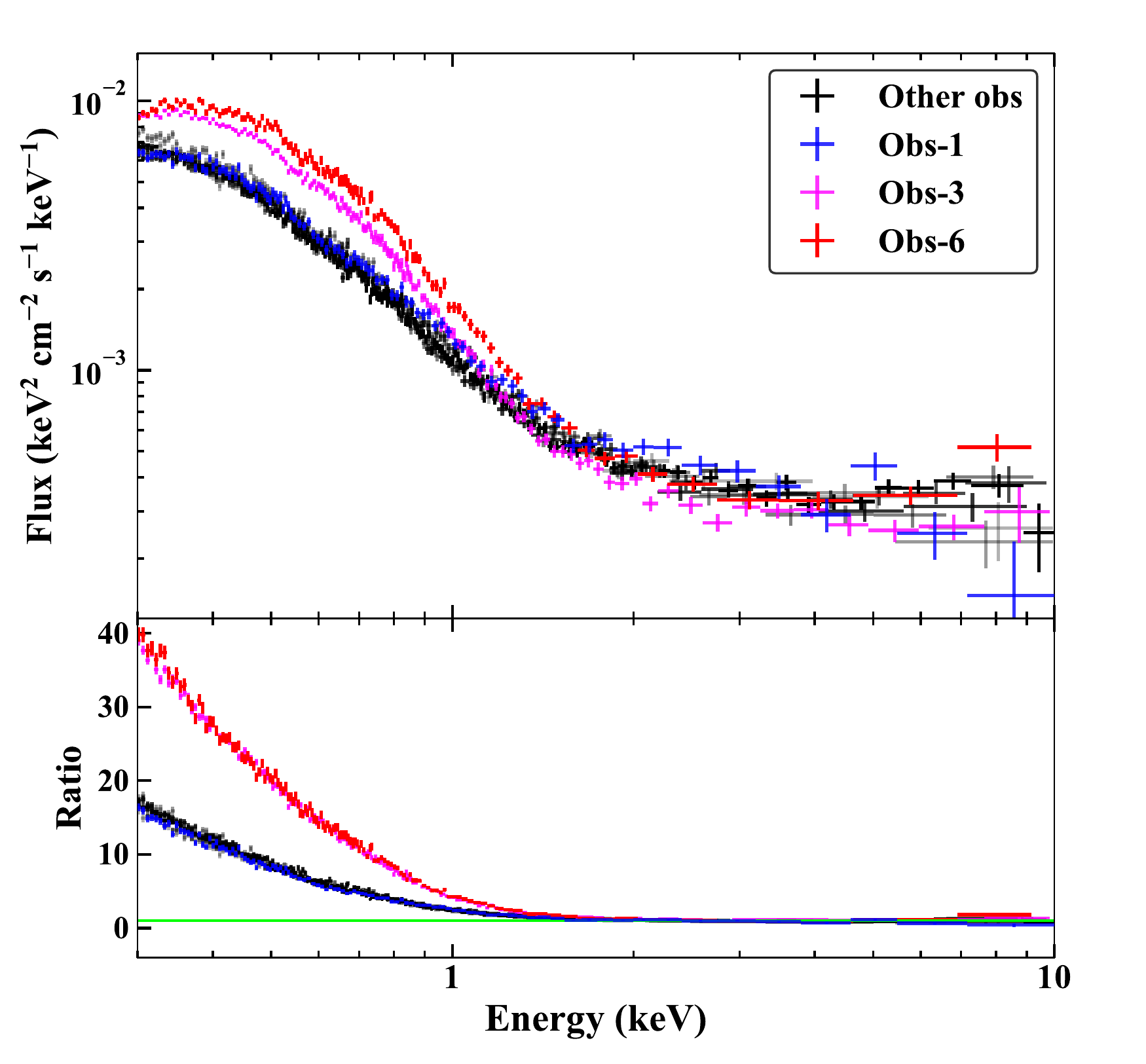}
\caption{The spectra of \rej1034\ in all the 9 \xmm\ observations. The blue, magenta and red spectra are from Obs-1, Obs-3 and Obs-6, respectively. Spectra from the other observations are presented in different grey scales as they have similar shapes. The ratio is relative to the best-fit hard X-ray power law component. It is clear that Obs-3 and Obs-6, which do not exhibit any QPO signal, have stronger soft X-ray excesses than the other observations with the QPO signal.}
\label{fig-allspec}
\end{figure}

\subsection{Anti-correlation between the QPO Detectability and the Soft X-ray Intensity}
\label{sec-spec-sx}
The trigger of the QPO in \rej1034\ is still not clear (Middleton et al. 2011). Previous studies have shown that the detection of this QPO is associated with the spectral hardness ratio, as the only two observations showing no QPO signal both have higher soft X-ray fluxes (Alston et al. 2014). To investigate this issue further, we perform simultaneous spectral fitting to all the time-averaged spectra from previous \xmm\ observations, with a typical spectral model of super-Eddington NLS1s which includes absorbed power law and a soft X-ray Comptonisation model (e.g. Jin et al. 2017a). The Galactic absorption is fixed at $N_{\rm H}=1.36\times10^{20}$ cm$^{-2}$ (Willingale et al. 2013), and the intrinsic absorption in \rej1034\ is left as a free parameter. This model fits the time-averaged spectra very well, with the total $\chi^2$ being 3942 for 3252 dof for all the 9 time-averaged EPIC-pn spectra. The slope of the soft excess is characterized by the photon index of a single power law fitted to the 0.3-2 keV spectrum.

The best-fit $N_{\rm H}$, fluxes and photon indices are all listed in Table~\ref{tab-obs}. The spectra and their ratios relative to the best-fit power law above 2 keV are shown in Figure~\ref{fig-allspec}. Note that Obs-1 is an exception because its clean exposure time is only 1.8 ks, and so we do not consider it as a useful dataset for the QPO study. In the rest observations, Obs-3 and Obs-6 have no QPO signal, and their soft excesses are much stronger and steeper than in other observations where a QPO can be detected. There is no significant difference in the hard X-ray flux or spectral slope between observations with and without a QPO, thus it is not very likely for the hard X-rays to contain the QPO trigger.

This anti-correlation between the QPO detectability and soft X-ray flux also appears consistent with the fact that the fractional rms variability of the QPO in 0.3-1 keV is less than that seen in harder X-rays. However, the soft X-ray fluxes of the two non-QPO observations are only a factor of 2-3 higher than the other QPO observations, so the non-detection of the QPO is not simply due to the dilution from an enhanced non-QPO soft X-ray component. We suggest that there should be some fundamental change in the accretion flow during Obs-3 and Obs-6, which enhances the soft X-ray emission and eliminates the QPO signal. This issue will be investigated in more detail in our Paper-II.

\section{Discussion}
\label{sec-qpo-origin}
The strong QPO signal in \rej1034\ is a rare phenomenon in AGN, and so its presence raises many interests and questions. Many models have been proposed to explain the QPO mechanism. For example, it was suggested that there is probably an X-ray emitting blob in the accretion flow of \rej1034\ which is periodically obscured by a warm absorber, so that the QPO signal is produced along with an absorption line whose variation is weakly correlated with the QPO's phase (Maitra \& Miller 2010, but also see Middleton, Uttley \& Done 2011). In order to explain the correlation between the instantaneous flux and QPO period, other models have been proposed, such as invoking a magnetic flare in a Keplerian orbit which has an intrinsic oscillation (Czerny et al. 2010), an oscillating shock in the accretion flow (Czerny et al. 2010; Das \& Czerny 2011; Hu et al. 2014), a spiral wave in a constant rotation state (Czerny et al. 2010), a temporary hot spot carried by the accretion flow with the Keplerian motion (Hu et al. 2014), the $g$-mode Diskoseismology caused by the gravitational-centrifugal force (Hu et al. 2014). However, due to the lack of more detailed characterisation of the QPO properties and its long-term variability, all these models remain poorly constrained. 

The new results concerning the QPO properties presented in this work have provided tests for some of these theoretical models. Firstly, we now  know that this particular QPO is a long-term, recurrent feature of this source, which appears in \rej1034\ from time to time during the past 11 years. This result suggests that the QPO is produced by a quite stable mechanism, and so disfavors models involving shorter timescales. For example, an X-ray emitting blob carried by the accretion flow at 10$R_{\rm g}$ away from a $\sim10^{6}M_{\sun}$ black hole would be accreted into the black hole within just a few months, but we observe a period shortening of only $250\pm100$ s over the past 11 years. So this QPO model can be ruled out with some confidence. 

Secondly, now that we have more observational information to examine which energy band produces the QPO. One of the main results is that in Obs-9 the QPO in 0.3-1 keV is leading 1-4 keV by $430\pm50$ s with a high coherence. This is more consistent with the possibility that the QPO is driven by a soft X-ray component, although the time lag alone is not sufficient to pin down the causality. This possibility is further supported by the anti-correlation found between the detectability of the QPO and the intensity of the soft X-ray excess, and also by the fact that the absolute rms amplitude of the QPO is larger in 0.3-1 keV than in harder X-ray bands. For comparison, we do not observe any systematic difference of the hard X-ray power law between the time-averaged spectra with and without the QPO (see Table~\ref{tab-obs}). For example, the hard X-ray photon indices and fluxes of the two non-QPO observations (i.e. Obs-3 and Obs-6) are not significantly different from the QPO observations. Therefore, it seems not likely that the origin of the QPO lies in the hard X-ray band. Moreover, the QPO frequency does not change significantly with the energy bands, so either the QPO-related soft and hard X-ray regions have similar sizes, or the mechanism is such that the QPO timescale does not depend on the size of its emitting region. One possibility is that the QPO arises from the soft X-ray band, and is transmitted to the hard X-ray band via Comptonisation of the QPO modulated soft emission.

Thirdly, the QPO in \rej1034\ has often been compared to the high-frequency QPOs in the micro-quasar GRS 1915+105. The similarity between \rej1034\ and GRS 1915+105 in terms of their X-ray spectra, PSD, and the super-Eddington accretion states suggest that the 67 Hz QPO in GRS 1915+105 may be an analogue of the QPO in \rej1034.
This was first proposed by Middleton et al. (2009) (also see Middleton, Uttley \& Done 2011; Done 2014), 
but the (apparent?) soft X-ray lag in the Obs-2 data is opposite to the soft X-ray lead seen in the 67 Hz high frequency QPO in GRS 1915+105 (M\'{e}ndez et al. 2013), thereby breaking the scaling relation. However, as we point out in this work, the 
associated coherence of the QPO time lag is low in 2007, so the soft lag is not very significant in these data, especially after
the pile-up correction. Instead, our new data from Obs-9 show that the highly coherent QPO in \rej1034\ has a soft X-ray lead, strongly supporting the analogy to the 67 Hz QPO in GRS 1915+105. Other features of the QPO are also consistent, e.g. small but significant changes in the 67~Hz QPO frequency are also seen in GRS~1915+105 (Belloni et al. 2019), similar to the fractional change in QPO frequency seen in \rej1034\ when comparing Obs-2 and Obs-9. The lack of a harmonic signal in \rej1034\ is not a concern, because the `harmonic' features in GRS 1915+105 do not appear simultaneously with the 67 Hz QPO very often, and they are all significantly weaker than the 67 Hz QPO (M\'{e}ndez et al. 2013).

So what then is the origin of the 67~Hz QPO in GRS 1915+105? These high frequency QPOs are rare in BHB, but are much more common in neutron-star X-ray binaries (XRB). In these objects we generally see two QPOs in the kHz region, an upper and a lower frequency separated by a few hundred Hz (see e.g. the review by van der Klis 2006). The lower frequency QPO shows a soft X-ray lag, while the upper frequency one generally shows a soft X-ray lead (de Avellar et al. 2013; Peille et al. 2015; Troyer et al. 2018). The (very rare) BHB high frequency QPOs are probably
the counterpart of the upper frequency QPO in neutron-star XRB (M\'{e}ndez et al. 2013). These are likely produced by 
oscillations within the Comptonising boundary layer between the disc and neutron star (Gilfanov et al. 2003; see also Karpouzas et al. 2020 for a detailed model of the lower frequency QPO). Whatever their origin is, we conclude that the QPO of \rej1034\ is indeed similar to the 67 Hz QPO in GRS 1915+105, where the soft X-rays lead the hard X-rays (M\'{e}ndez et al. 2013).

\section{Conclusions}
\label{sec-conclusion}
In this paper we report the detection of a strong X-ray QPO in the new \xmm\ observation of \rej1034\ in 2018, which is separated by 7 years from the previous \xmm\ observation and 11 years from the original discovery of this QPO signal. New and detailed analysis have been conducted that verify and extend the QPO properties previously known, which are summarized below:

\begin{itemize}
\item we confirm that the X-ray QPO in \rej1034\ is a robust phenomenon which has occurred, at least intermittently, for more than 11 years. Its presence is most significant in the latest observation taken in 2018, which yields a 9$\sigma$ significance of detection in the 1-4 keV band. The quality value is $\sim$20, and the folded light curve exhibits a well defined sinusoidal shape, and so the QPO is highly coherent. 
\item in the new Obs-9 data the QPO period is found to be $3530\pm80$ s in the 1-4 keV band, and shows no significant change with energy bands. However, the fractional rms of the QPO increases from 4\% in 0.3-1 keV to 12.4\% in the 1-4 keV band, although the absolute rms amplitude of the QPO in 0.3-1 keV is actually a factor of 2.4 higher than in the 1-4 keV band.
\item we find that the QPO period is shorter in the new observation than was observed before. It was $3800\pm70$ s in the 1-4 keV light curve in Obs-2, but decreases by $250\pm100$ s in Obs-9 (i.e. $\sim$7\% of the QPO period). The significance of this long-term variation of the QPO period is also confirmed by our simulations performed following the Bayesian approach.
\item our analysis shows that the QPO the 0.3-1 keV band leads the 1-4 keV band by $430\pm50$ s, and the time lag is accompanied by a high coherence. This result is further confirmed by the direct folding of the light curves in these energy bands. This soft X-ray lead is opposite to the soft X-ray lag reported previously for Obs-2. However,our re-analysis of these data indicates that the QPO lag found in Obs-2 is associated with a lower coherence, and so it is less robust than that observed in Obs-9. Therefore, either the previously reported soft X-ray lag is not intrinsic to the QPO, or the lag has changed direction from Obs-2 to Obs-9, which if confirmed would be an interesting new phenomenon to explain. Clearly future observations are required to address this issue.
\item by analyzing the data from all previous \xmm\ observations, we show that the two observations without a QPO show stronger soft X-ray excesses than the other observations which display evidence of a QPO. Therefore, we conclude that there is a long-term anti-correlation between the intensity of the soft X-ray excess and the detectability of a QPO signal. 
\end{itemize}

These newly discovered and refined properties of the QPO in \rej1034\ show that it is more similar to the 67 Hz QPO in \grs1915. We also suggest that the QPO of \rej1034\ is probably driven by a soft X-ray component. In order to further understand the mechanisms of the QPO and the soft excess, we will present a more comprehensive spectral-timing analysis for the QPO together with broader frequency ranges in our forthcoming Paper-II. Finally, we emphasize the importance of carrying out long-term monitoring of the QPO and the spectral state of \rej1034. This source provides one of the best laboratories in which to study the physics of the QPO phenomenon in AGN, and so we recommend it to be one of the highest-priority AGN targets for future X-ray missions such as the Einstein Probe mission ({\it EP}), the enhanced X-ray Timing and Polarimetry mission ({\it eXTP}) and the Advanced Telescope for High-ENergy Astrophysics ({\it Athena}).

\section*{Acknowledgements}
We thank the anonymous referee for providing useful comments to improve the paper.
CJ acknowledges the National Natural Science Foundation of China through grant 11873054, and the support by the Strategic Pioneer Program on Space Science, Chinese Academy of Sciences through grant XDA15052100.
CD and MJW acknowledge the Science and Technology Facilities
Council (STFC) through grant ST/P000541/1 for support.
This work is based on observations conducted by \xmm, an ESA
science mission with instruments and contributions directly funded by
ESA Member States and the USA (NASA).
This research has made use of the NASA/IPAC Extragalactic Database (NED) which is operated by the Jet Propulsion Laboratory, California Institute of Technology, under contract with the National Aeronautics and Space Administration.











\bsp	
\label{lastpage}
\end{document}